\documentclass[journal]{IEEEtran}
\ifCLASSINFOpdf
\else
\fi
\usepackage[utf8]{inputenc} % allow utf-8 input
\usepackage[T1]{fontenc}    % use 8-bit T1 fonts
\usepackage{hyperref}       % hyperlinks
\usepackage{url}            % simple URL typesetting
\usepackage{booktabs}       % professional-quality tables
\usepackage{threeparttable}
\usepackage{diagbox}
\usepackage{amsfonts}       % blackboard math symbols
\usepackage{nicefrac}       % compact symbols for 1/2, etc.
\usepackage{microtype}      % microtypography
\usepackage{lipsum}
\usepackage{color}
\usepackage{fancyhdr}       % header
\usepackage{graphicx}       % graphics
\graphicspath{{media/}}     % organize your images and other figures

\begin{document}
%
% paper title
% Titles are generally capitalized except for words such as a, an, and, as,
% at, but, by, for, in, nor, of, on, or, the, to and up, which are usually
% not capitalized unless they are the first or last word of the title.
% Linebreaks \\ can be used within to get better formatting as desired.
% Do not put math or special symbols in the title.
%% Title
\title{Synthetic Datasets for Autonomous Driving: A Survey
%%%% Cite as
%%%% Update your official citation here when published 
}

\author{Zhihang~Song, 
	Zimin~He,
        Xingyu~Li, 
        Qiming~Ma,
        Ruibo~Ming,
        Zhiqi~Mao,
        Huaxin~Pei,
        Lihui~Peng,~\IEEEmembership{Senior~Member,~IEEE},
        Jianming~Hu,~\IEEEmembership{Member,~IEEE},
        Danya~Yao,~\IEEEmembership{Member,~IEEE},
	and~Yi~Zhang,~\IEEEmembership{Member,~IEEE}

\thanks{This work was supported by the National
Natural Science Foundation of China under Grant 62133002. (\textit{Zhihang Song and Zimin He contributed equally to this work.}) \ (\textit{Corresponding author: Lihui Peng.})}
\thanks{Zhihang Song, Zimin He, Xingyu Li, Qiming Ma, Ruibo Ming, Zhiqi Mao, and Huaxin Pei are with the Department of
Automation, Tsinghua University, Beijing 100084, China (e-mail: \{song-zh22, hzm22, lixingyu22, mqm22, mrb22\}@mails.tsinghua.edu.cn; \{mzq16, phx17\}@tsinghua.org.cn).}
\thanks{Lihui Peng is with the Department of Automation, Institute of Measurement and Applied Electronics, Tsinghua University, Beijing 100084, China. (e-mail: lihuipeng@mail.tsinghua.edu.cn).}
\thanks{Jianming Hu, Danya Yao, and Yi Zhang are with the Department of Automation, BNRist, Tsinghua University, Beijing 100084, China, and also with the Jiangsu Province Collaborative Innovation Center of Modern Urban Traffic Technologies, Nanjing 210096, China (e-mail: \{hujm, yaody, zhyi\}@mail.tsinghua.edu.cn).}}

\maketitle

\begin{abstract}
Autonomous driving techniques have been flourishing in recent years while thirsting for huge amounts of high-quality data. However, it is difficult for real-world datasets to keep up with the pace of changing requirements due to their expensive and time-consuming experimental and labeling costs. Therefore, more and more researchers are turning to synthetic datasets to easily generate rich and changeable data as an effective complement to the real world and to improve the performance of algorithms. In this paper, we summarize the evolution of synthetic dataset generation methods and review the work to date in synthetic datasets related to single and multi-task categories for the autonomous driving perception study. We also discuss the role that synthetic datasets play in the evaluation, gap test, and positive effect of autonomous driving-related algorithm testing, especially on trustworthiness and safety aspects, and provide examples of evaluation experiments. Finally, we discuss the limitations and future directions of synthetic datasets. To the best of our knowledge, this is the first survey focusing on the application of synthetic datasets in autonomous driving. This survey also raises awareness of the problems of real-world deployment of autonomous driving technology and provides researchers with a possible solution.
\end{abstract}

% keywords can be removed

\section{Introduction}
Recent trends in autonomous driving have led to a proliferation of studies in perception and system testing algorithms, which are thirsty for high-quality data. A large amount of rich and accurate data is not only the basis of algorithm experiments but also affects algorithm performance. {To meet the needs of various autonomous driving perception tasks, such as object detection and tracking, instance segmentation, depth estimation, optical flow estimation, etc., KITTI \cite{doi:10.1177/0278364913491297}, Waymo Open \cite{sun2020scalability}, BDD100K \cite{Yu2018BDD100KAD}, and more other large-scale real-world datasets have been created to provide an experimental basis and platform for the development of autonomous driving algorithms. These datasets are now widely used as benchmark datasets for training and testing perception algorithms.}

However, datasets from real scenes and sensors also have certain limitations and drawbacks due to their physical experimental nature. Such experiments often require expensive hardware (such as vehicles, LiDARs, etc.) and a lot of labeling work. For more complex problems, such as studying the effect of certain variables on perception tasks, more experiment time is often needed, because some factors, such as weather and corner cases, are not controllable, limiting more detailed research in these rare but important cases. In addition, when system testing has specific requirements for the collected data, such as extending the coverage and variety of scenarios and weather, much longer driving distances and times are required, reducing the speed and reliability of testing. Some researchers have found that one of the most effective methods to overcome the limitations of real-world data is to create synthetic datasets \cite{he2022synthetic, paulin2023review, nikolenko2021synthetic}. This effectiveness lies in the ability to carefully control and manipulate various environmental factors, scenarios, and data distributions that are difficult or even impossible to achieve in real-world data collection.

\begin{figure*}[htbp]
	\centerline{\includegraphics[width=0.8\linewidth]{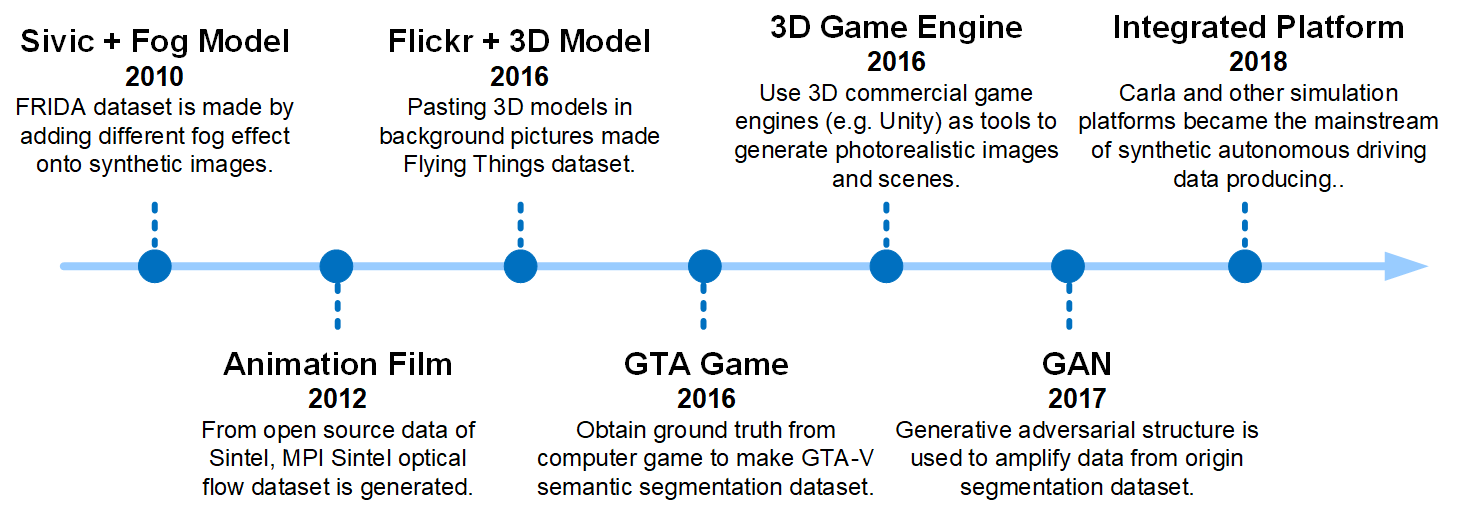}}
	\caption{Different development stages of synthetic dataset generation for autonomous driving perception related tasks.}
	\label{fig:fig1}
\end{figure*}

{The inception of synthetic data can be traced back to the 1960s, used to generate simpler line drawings for computer vision algorithms than real data \cite{nikolenko2021synthetic, huffman1971impossible}. Then synthetic data was used as a test set for fixed algorithms like the Lucas-Kanade optical estimation method \cite{lucas1981iterative}. With the development of deep learning, not only limited to the widely known computer vision field, synthetic data is also used as a training source in neural programming \cite{gers2001lstm}, bioinformatics \cite{schneider2016novo}, and neural language processing \cite{nikolenko2021synthetic, fadaee2017data}.} With regard to the techniques for generating synthetic datasets related to autonomous driving, they have gone through several stages of development. See Fig.~\ref{fig:fig1}. 

{The first publicly available synthetic dataset applied to autonomous driving tasks is the Fog Road Image Database (FRIDA) generated by Tarel et al. \cite{jpt-iv10} using Sivic software in 2010. Aimed at solving the depth estimation task in foggy weather, FRIDA provides 90 synthetic images of 18 urban road scenes with different types of synthetic fog. And FRIDA2 was published in 2012 with more images and road scenes \cite{jpt-itsm12}.} In 2012, Max Planck Institute (MPI) Sintel, generated by Bulter et al. \cite{Butler2012ANO} using the animated movie Sintel as a data source, the MPI Sintel dataset contains 1,041 pairs of images for optical flow estimation and has become the first widely used synthetic dataset. However, the accessibility of data from open-source movies is relatively limited, mainly due to the fact that only a small number of movies have adopted licensing agreements such as the \textit{Creative Commons Attribution 3.0 License} \cite{CC3} or similar permissions that allow researchers to freely use the content to build datasets.

In 2016, Mayer et al. \cite{mayer2016large} utilized 3D object models from the ShapeNet database \cite{7301289} to create Flying Things optical flow dataset with about 25,000 stereo frames and built the Driving dataset with the same model pool but a more naturalistic street scene resembling the KITTI dataset. But the method of splicing images and 3D models suffers from low fidelity and relatively simple scenes. Therefore, researchers have gradually explored various methods to synthesize autonomous driving datasets using existing image resources. In 2016, Richter et al. \cite{101007} developed a tool to generate pixel-level semantic segmentation annotations from modern computer games without requiring open-source code. This method can quickly complete semantic annotations from the communication between games and graphics hardware. Virtual KITTI was also launched this year by Gaidon et al. \cite{Gaidon2016VirtualWorldsAP}. {Since then, the commercial 3D game engines (such as Unity) have been consistently used as the generation tool and have proven to be a popular choice for creating synthetic datasets.}

In 2017, Isola et al. \cite{Isola2016ImagetoImageTW} generated larger semantic segmentation data from original data with the Pix2Pix network of generative adversarial structure. Further work on Pix2PixHD was done by Wang et al. in 2018 \cite{Wang2017HighResolutionIS}, which can add more details to synthesize scene images from semantic annotations. In addition, CRN \cite{Chen2017PhotographicIS}, SIMS \cite{Qi2018SemiParametricIS}, GauGAN \cite{Park2019SemanticIS}, SelectionGAN \cite{Tang2019MultiChannelAS}, TSIT \cite{Jiang2020TSITAS}, PIS \cite{Dundar2020PanopticBasedIS}, DAGAN \cite{Tang2020DualAG}, SMIS \cite{Zhu2020SemanticallyMI}, MaskGAN \cite{Lee2019MaskGANTD}, and SEAN \cite{Zhu2019SEANIS} are also classic algorithms for synthesizing scene images from semantic images using GAN structure. {As demonstrated in \cite{veeravasarapu2017adversarially}, GANs were initially used to generate 3D scene layouts. And recently it has also been used with structures such as U-Net to generate synthetic images \cite{li2023novel}.} In their early application to image generation, it is challenging to have fine-grained control over the generated output, and sometimes there exist deformation and abnormal detailed structures, which has a negative impact on the feasibility of automating the annotation process.

In 2018, the launch of the Carla platform \cite{Dosovitskiy2017CARLAAO}, based on the Unreal Engine 4 (UE4), provided a novel systematic solution for performance testing and dataset generation in autonomous driving scenarios. Since then, more datasets have been conveniently synthesized through the Carla simulation platform and its supporting sensor suites due to the realistic, flexible, and open-source usage mode. Nowadays, with the development of large language models (LLM), it is also combined with LLM to simulate various and controlled environments \cite{tian2023vistagpt}.

The proliferation of synthetic datasets has prompted researchers to conduct surveys on this topic \cite{paulin2023review, nikolenko2021synthetic}. While these surveys cover various methods and techniques for generating synthetic datasets, they often overlook the specific synthetic datasets related to autonomous driving tasks. This gap motivates us to delve into the study of synthetic datasets in the context of autonomous driving. In this study, we select a subset of synthetic datasets based on their widespread recognition, extensive adoption, relevance to autonomous driving perception tasks, and availability of ground truth annotations. Furthermore, these datasets are selected for their diversity in representing different driving scenarios. They have played a central role in previous research and are critical for in-depth analysis of autonomous driving development and evaluation. Our goal is to highlight the autonomous driving perception tasks included in these synthetic datasets, explore the challenges associated with generating synthetic datasets, and thus provide guidance for the evaluation and generation.

In the early stages, synthetic datasets primarily focused on a particular autonomous driving perception task, such as optical flow estimation or semantic segmentation. However, as the generation techniques advanced, the synthetic datasets became more diverse and applicable to a broader range of tasks. These tasks include 2D/3D object detection, multi-object tracking, depth estimation, and various others. Consequently, we can categorize the existing synthetic datasets related to autonomous driving into two main groups: single-task datasets and multi-task datasets, based on the types of tasks they encompass.

The main contributions of this survey include: (a) We present a comprehensive survey of synthetic datasets designed for {perception} tasks related to autonomous driving, providing detailed insights into both single-task and multi-task synthetic datasets. To the best of our knowledge, this is the first survey that focuses on synthetic datasets utilized in the field of autonomous driving. (b) We propose a framework for evaluating synthetic datasets to facilitate the generation of trustworthy datasets. This framework serves as a valuable foundation for future research efforts in the areas of synthetic dataset evaluation and generation. (c) We propose a systematic process for generating synthetic datasets with evaluation feedback for autonomous driving. This process ensures the production of trustworthy datasets that can be effectively evaluated and iteratively improved to meet the specific requirements of autonomous driving research and development.

To better present our work, the rest of this paper is organized as follows. Section \ref{Motivation} describes the motivation for generating synthetic datasets in the field of autonomous driving. Sections \ref{Single-Task} and \ref{Multi-Task} introduce the single-task and multi-task synthetic datasets in detail, respectively. In Section \ref{Evaluation}, we propose an evaluation framework for synthetic datasets and demonstrate how they contribute to safe and trustworthy autonomous driving systems. In Section \ref{Experiments}, we conduct experiments of the proposed evaluation method on dataset examples. Finally, we discuss some perspectives for future research in Section \ref{Future} and provide a summary in Section \ref{Conclusion}.

\section{Motivation} \label{Motivation}
Previous studies have emphasized the significance of data in the development of autonomous driving systems. However, keeping up with the rapid evolution of emerging algorithms using large real-world datasets can be challenging. {Synthetic data have attracted increasing attention from researchers due to their cost-effectiveness and speed of acquisition,} and synthetic datasets do exhibit some advantages over real-world datasets. In this section, we outline several reasons why synthetic datasets are becoming more widely used.

First, building large-scale autonomous driving datasets is often costly in the real world. Obtaining sufficient quantities of high-quality raw data requires specific sites and many experiments. Also, the precise manual labeling of multi-modal sensors or different tasks requires hard and long-term work, often requiring expensive assistance from a professional team. For example, labeling one segment by an individual worker averages 0.86 USD, and one image often has 15 to 30 segments \cite{kim2020reducing}. Such a high cost imposes many restrictions on the development of real datasets, especially the long update and iteration cycles.

Second, for research on some specific issues, such as exploring the performance of algorithm models under different domain shifts (such as changes in weather and lighting conditions), it is difficult to obtain various controllable condition changes in real scenarios. Consequently, researchers often rely on synthetic data as a valuable resource to overcome these limitations and conduct comprehensive studies.  

Third, due to the serious long-tail problem of real datasets in system safety testing, most of the collected data comes from routine driving scenarios. Thus, the occurrence frequency of challenging scenarios (such as traffic accidents, illegal driving, etc.) that are critical to safety is too low. It becomes necessary to identify \cite{song2023identifying}, design and execute numerous test cases to ensure the robustness of the system, which significantly increases the cost of testing. In contrast, synthetic datasets can formulate safety-relevant scenarios according to specific requirements, improving the speed and reliability of testing. For example, scenarios engineering (SE) is already being used to generate challenging synthetic scenarios \cite{guo2023vectorized}.

Since the synthetic autonomous driving dataset offers the advantages of ease of acquisition and flexibility in generation, it represents a potential solution to the aforementioned challenges. {It is important to note that although the initial cost of generating synthetic data may be relatively higher compared to capturing and annotating a single real image, the cost-effectiveness tends to improve as the volume of synthetic data increases.} This type of data generation can automatically generate accurate labels without manual effort, making it convenient to efficiently update and iterate datasets, and can also exhibit rich specificity as requirements change, reducing the impact of overfitting problems. Researchers hope to use computer-generated virtual scene data as a substitute or supplement for the real world so as to improve the performance of algorithms and systems in the real world. {At the same time, the emergence of the concept of trustworthy artificial intelligence (AI) emphasizes the importance of evaluating and testing the credibility of algorithms, especially those related to machine learning and AI models. This aspect is also closely related to the design and application of synthetic datasets.}

\section{Single-Task Synthetic Datasets} \label{Single-Task}
\begin{figure}[htbp]
\centerline{\includegraphics[width=1\linewidth]{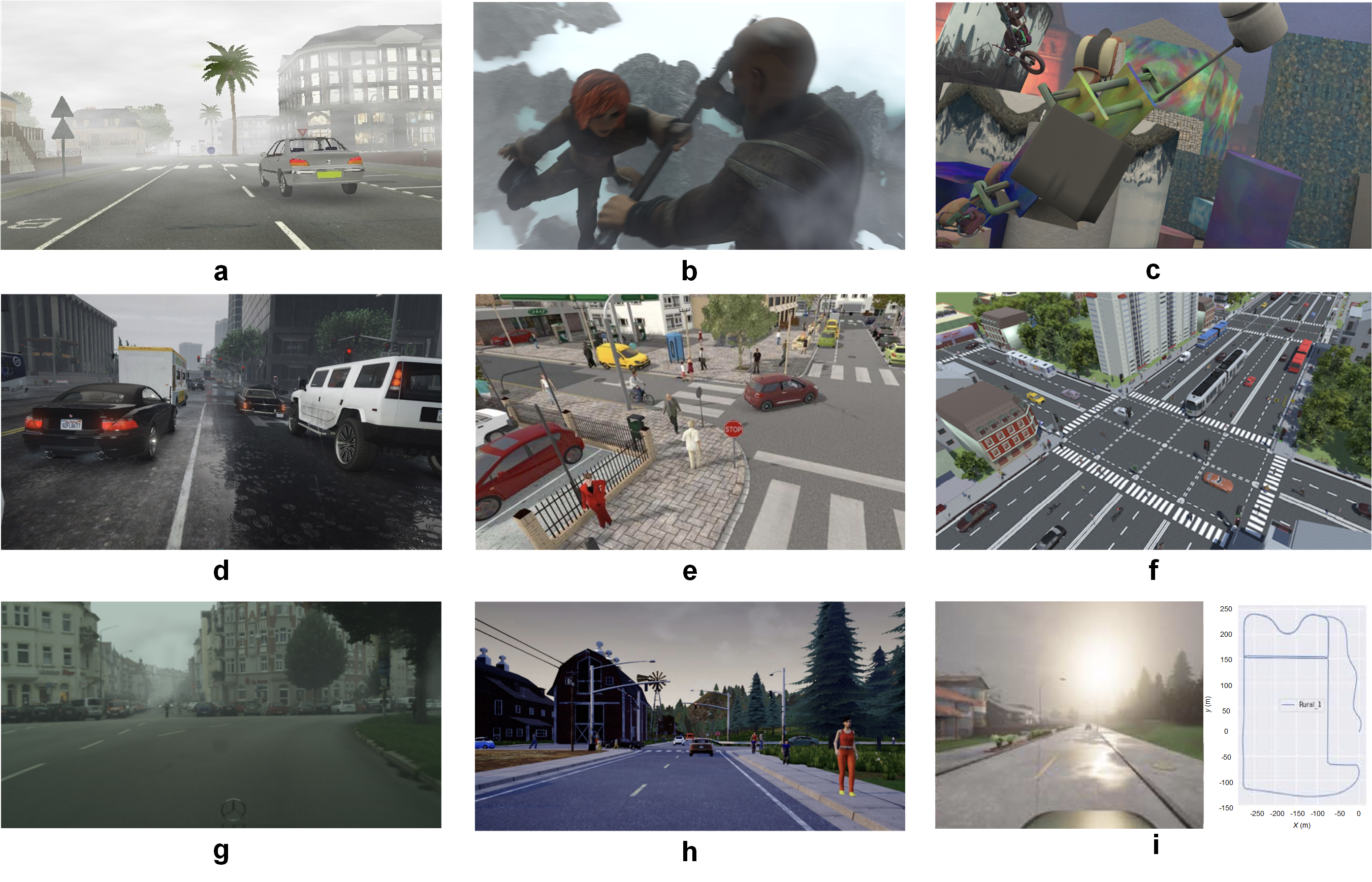}}
\caption{Samples of single-task synthetic datasets. a) FRIDA, b) MPI Sintel, c) Flying Things, d) GTA-V {dataset}, e)
SYNTHIA, f) VEIS, g) Foggy Cityscapes, h) IDDA, i) CarlaScenes.}
\label{fig:single-task}
\end{figure}

Early synthetic datasets focused on single tasks, such as depth estimation, optical flow estimation, and semantic segmentation, which significantly reduced the reliance on manual data annotation. These datasets paved the way for the creation of multi-task synthetic datasets. In this section, we provide a brief overview and performance evaluation of these datasets, samples of which are shown in Fig.~\ref{fig:single-task}, and the comparison is shown in TABLE~\ref{tab1}.

\begin{table*}[htbp]
\begin{center}
\caption{Comparison of Single-task Synthetic Datasets for Autonomous Driving}
\label{tab1}

\begin{threeparttable}
\begin{tabular}{l*{14}{c}}
\midrule

\textbf{\textit{Dataset}} & 
\textbf{\textit{Frames}}
&\textbf{\textit{Generation Tools}}& \textbf{\textit{Task}}&\\
\midrule
{FRIDA} \cite{jpt-iv10}&{90}&Sivic&{Depth estimation}&\\
MPI Sintel \cite{Butler2012ANO}&1K&{Blender\tnote{1}}&Optical flow estimation&\\
Flying Things \cite{mayer2016large}&25K&{Blender\tnote{1}}&Optical flow estimation&\\
GTA-V {dataset} \cite{101007}&25K&GTA game&Semantic segmentation&\\
SYNTHIA \cite{Ros2016TheSD}&213K&{Unity development platform}&Semantic segmentation&\\
VEIS \cite{saleh2018effective}&61K&{Unity development platform}&Instance segmentation&\\
Foggy Cityscapes \cite{Sakaridis2017SemanticFS}&20K&MATLAB platform&Semantic segmentation&\\
IDDA \cite{Alberti2020IDDAAL}&1M&{Carla platform}&Semantic segmentation&\\
CarlaScenes \cite{Kloukiniotis2022CarlaScenesAS}&-&{Carla platform}&Odometry&\\
\bottomrule
\end{tabular}
\begin{tablenotes}
    \footnotesize 
    \item[1] https://www.blender.org/
\end{tablenotes}     
\end{threeparttable}
\end{center}
\end{table*}

\subsection{FRIDA} 
{FRIDA \cite{jpt-iv10}, published by Tarel et al. in 2010, is a synthetic dataset with a focus on fog conditions. They constructed a realistic 3D urban environment using Sivic software and applied Koschmieder’s law to simulate various types and densities of fog. Each image without fog is paired with 4 foggy versions and a depth map. FRIDA consists of 90 synthetic images of 18 urban road scenes and provides a valuable resource for researchers to explore the relationship between fog density and scene depth.} 

{Building on FRIDA, researchers released an expanded dataset called FRIDA2 in 2012 \cite{jpt-itsm12}. FRIDA2 contains 330 synthetic images of 66 different road scenes. The presence of both clear and foggy images in FRIDA and FRIDA2 also facilitates the evaluation of visibility enhancement and dehazing algorithms \cite{agrawal2022comprehensive, guo2023haze}. These algorithms are critical components for improving the performance of autonomous driving systems in challenging weather conditions.}

\subsection{MPI Sintel} 
MPI Sintel \cite{Butler2012ANO}, created by Bulter et al. in 2012, was specifically designed for optical flow estimation. Optical flow estimation helps vehicles perceive the temporal continuity of the environment and plays an important role in time-series tasks \cite{shen2023optical}. For example, vehicle sensors can recover the 3D structure and motion of objects from the optical flow to provide reliable support for more advanced vision tasks, ultimately controlling vehicle steering and speed \cite{kondermann2016hci}.

This dataset is based on the open-source 3D animated short film Sintel and contains 1,064 synthetic stereo images and real-world data from 23 different scenes. The generation process is described in detail in \cite{Wulff:ECCVws:2012}. The statistical properties of MPI Sintel were analyzed and found to be similar to those of natural movies. However, MPI Sintel often ignores physical constraints, and caution should be exercised when using this dataset to train and evaluate algorithms that rely heavily on real-world physical laws \cite{Butler2012ANO}. 

To facilitate the validation of the optical flow estimation model, a publicly accessible evaluation website has also been created, which demonstrates several methods dedicated to optical flow estimation, such as PWC-Net \cite{sun2018pwc}, SelFlow \cite{liu2019selflow}, and ScopeFlow \cite{bar2020scopeflow}. To date, MPI Sintel has become one of the most widely used datasets for optical flow tasks because it represents natural scenes and motion well.

\subsection{Flying Things} 
Flying Things \cite{mayer2016large} was introduced by Mayer et al. in 2016. It extends the previous optical flow estimation task to include disparity estimation and scene flow estimation. The generated motion cut-and-paste foreground is used to obtain training data pairs. This dataset focuses on a collection of daily objects flying along random 3D trajectories and is generated using randomness and advanced rendering capabilities. It consists of over 25,000 stereo image pairs, including ground truth data for disparity, optical flow, and scene flow.

The simulation results demonstrate that Flying Things can successfully train large convolutional networks and support challenging tasks with high accuracy. Other optical flow synthetic datasets have been created using the same method as Flying Things. For example, the driving dataset \cite{mayer2016large} utilizes the same vehicle model as Flying Things, with a naturalistic dynamic street scene and a viewpoint similar to the KITTI dataset.

\subsection{GTA-V {dataset}} 
Grand Theft Auto V (GTA-V) {dataset} \cite{101007}, published by Richter et al. in 2016, is a pixel-level semantic segmentation dataset that relies on the realism of commercial video games. It features accurate simulations of material appearance, light transmission, player interaction, and realistic placement of objects and environments. Specifically, it incorporates a wrapper between the game and the operating system to facilitate the capture, modification, and reproduction of rendering commands. By hashing various rendering resources such as geometry, textures, and shaders that are passed from the game to the graphics hardware, it generates object signatures across scenes and game sessions. This process allows for the creation of pixel-level object labels without the need for boundary tracking.

By incorporating GTA-V {dataset} into existing real-world datasets such as CamVid and KITTI, model training accuracy can be greatly improved while minimizing the need for expensive manual annotation of real-world data \cite{101007}. Some state-of-the-art models, such as MIC \cite{hoyer2022mic}, HRDA \cite{hoyer2022hrda}, and SePiCo \cite{xie2023sepico}, have also shown great performance in the semantic segmentation of GTA-V {dataset}.

\subsection{SYNTHIA} 
SYNTHIA \cite{Ros2016TheSD} was released by Ros et al. in 2016 to address the challenge of semantic segmentation and improve the understanding of scenes related to driving. This dataset is generated by the virtual city rendering in the Unity development platform with appropriate material coefficients to ensure realistic effects and different seasons with variations in appearance. In addition, the dynamic lighting engine can create different lighting conditions to simulate changes in daylight. With over 213,400 synthetic images captured from different viewpoints, seasons, weather conditions, and lighting, it provides pixel-level annotations for 13 categories, including sky, building, road, sidewalk, fence, vegetation, lane-marking, pole, car, traffic signs, pedestrians, cyclists, and miscellaneous. Simulation results show that the models trained on SYNTHIA and fine-tuned on the real-world dataset achieve improved semantic accuracy.

In addition, SYNTHIA-San Francisco (SYNTHIA-SF) \cite{Jurez2017SlantedSR} and SYNTHIA-AL \cite{Bengar2019TemporalCF} have also been released on the basis of SYNTHIA. SYNTHIA-SF contains 2,224 synthetic images with accurate depth information and 19 classes of semantic annotations, making it ideal for evaluating the accuracy of depth and semantic segmentation. SYNTHIA-AL, on the other hand, includes annotations for instance segmentation, 2D and 3D bounding boxes, and depth information, specifically designed for evaluating active learning in road scenes for video target detection.

\subsection{VEIS}
Virtual Environment for Instance Segmentation (VEIS) \cite{saleh2018effective}, created by Saleh et al. in 2018, provides a synthetic environment for instance segmentation tasks. Built on the Unity3D game engine, VEIS employs a virtual camera mounted on a virtual vehicle to capture urban environments. Each instance is assigned a unique ID, and the system renders textures and shaders while simultaneously capturing synthetic images and instance-level semantic segmentation maps in real time. The dataset contains 61,305 frames of multi-class and single-class scenes, annotated with instance segmentation of the foreground class.

A framework for training instance segmentation on purely synthetic data is proposed, where background and foreground are processed independently and the segmentation results are then combined. This framework helps autonomous vehicles effectively discriminate between foreground and background classes, enabling efficient instance segmentation.

\subsection{Foggy Cityscapes}
Foggy Cityscapes \cite{Sakaridis2017SemanticFS}, developed by Sakaridis et al. in 2018, is a synthetic semantic segmentation dataset characterized by images with fog. By implementing an optical model of fog in the MATLAB platform, synthetic fog is added to the Cityscapes dataset \cite{Cordts2016Cityscapes} to create images with fog. It consists of 20,550 images, each annotated with fine-grained semantic annotations from Cityscapes. The fog in the images was created by synthesizing 20,000 images from a larger, coarsely annotated dataset and 550 high-quality images.

Foggy Driving was created to evaluate model performance in real-world fog scenarios. Research has shown that Foggy Cityscapes could improve model performance for semantic segmentation and target detection in challenging foggy environments, in both fully and semi-supervised settings. This dataset can also be applied to real-time rapid defogging systems \cite{mandal2020real}.

\subsection{IDDA}
ItalDesign DAtaset (IDDA) \cite{Alberti2020IDDAAL} was created by Alberti et al. in 2020 and is the largest synthetic semantic segmentation dataset available. On the Carla platform, an RGB camera, a semantic segmentation sensor, and a depth sensor are used to generate the dataset. Using object blueprints in the UE4, the semantic segmentation sensor can generate pixel-level labeled images in real time, while the depth sensor captures images that encode depth information in RGB color space. The dataset contains over 1 million images, with more than 100 different combinations of scenes, 5 perspectives, 7 cities, and 3 weather conditions. The images are annotated with pixel-level semantic information and depth maps.

Comparisons were made between semantic segmentation methods with and without domain adaptation, demonstrating that IDDA serves as an effective evaluation benchmark for assessing the performance of these methods in domain transfer. The network trained on IDDA was also applied to a real-world dataset, revealing that while there remains a notable performance gap in domain adaptation techniques, IDDA exhibits a certain degree of proximity to real-world scenarios. Moreover, the feature embedding method was employed to quantify the disparity between IDDA and real-world datasets \cite{gadipudi2022synthetic}, presenting valuable implications for further research in the field of domain adaptation.

\subsection{CarlaScenes}
CarlaScenes \cite{Kloukiniotis2022CarlaScenesAS} was developed by Kloukiniotis et al. in 2022 to measure the mileage of autonomous vehicles. It was generated using the Carla platform and includes seven difficult-to-measure scenarios, such as uphill and downhill, rural environments, circular roads, different weather and lighting conditions, multiple moving objects, complex urban environments, and long-distance roads. The ego vehicle travels through these scenarios, and data is collected by cameras, LiDARs, IMU, and GNSS for the odometry measurement.

Three monocular vision methods (DSO \cite{engel2017direct}, LEGO LOAM \cite{shan2018lego}, and DVSO \cite{yang2018deep}) were tested on the dataset and found to be unsuitable for these challenging scenarios. However, it is believed that with proper training and validation, deep learning-based odometry methods can accurately measure a vehicle's mileage in various scenarios. CarlaScenes provides a convenient solution to the difficult task of tagging real-world scenario data.

\section{Multi-task Synthetic Datasets} \label{Multi-Task}
\begin{figure}[htbp]
\centerline{\includegraphics[width=1\linewidth]{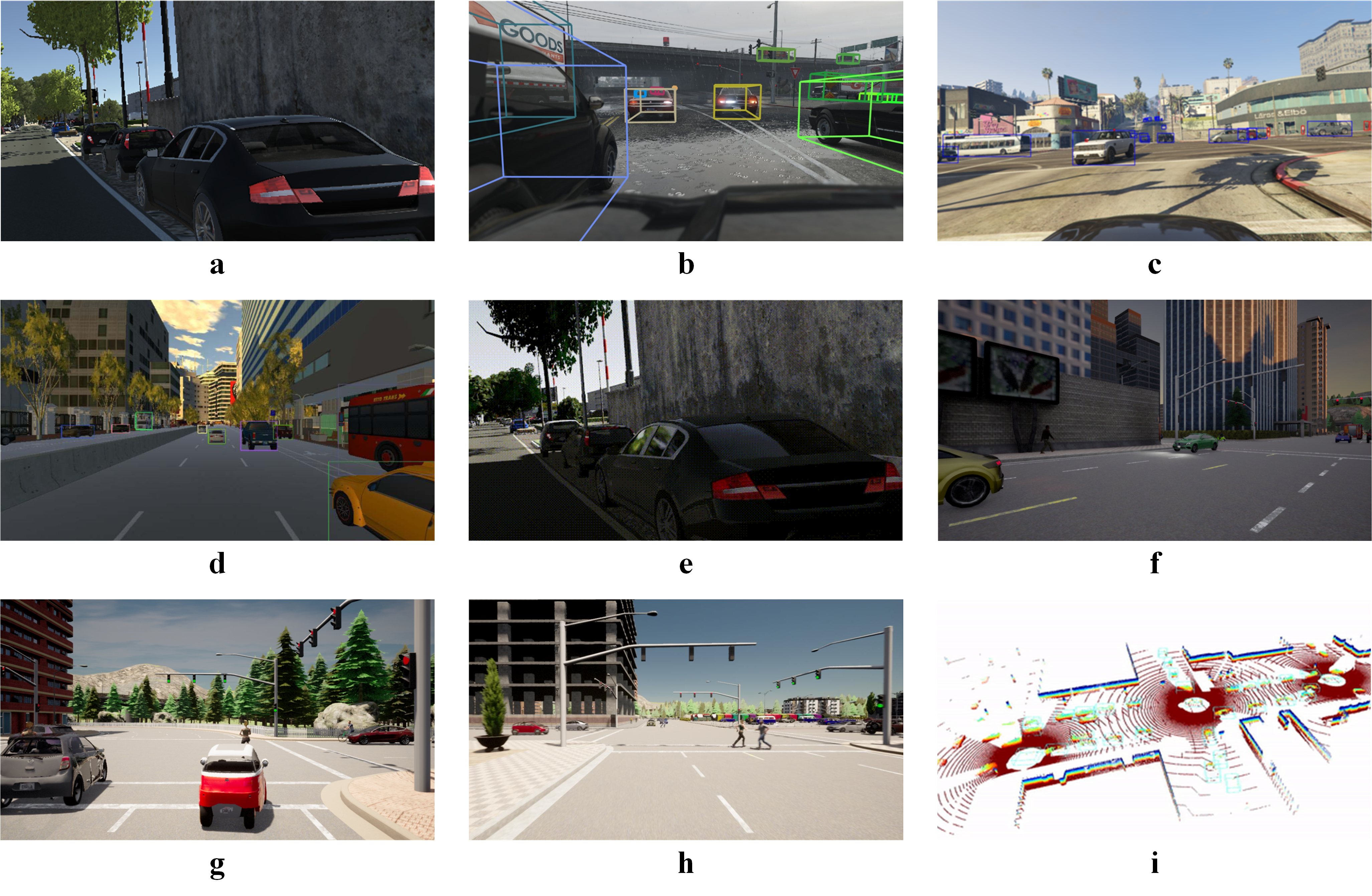}}
\caption{Samples of multi-task synthetic datasets. a) Virtual KITTI, b) VIPER, c) ParallelEye, d) PreSIL, e) Virtual KITTI2, f) SHIFT, g) V2X-Sim, h) AIODrive, i) OPV2V.}
\label{fig:multi-task}
\end{figure}

In order to solve the problems of expensive labeling costs and limited test scenarios faced by real-world datasets, relying on increasingly mature virtual engines such as Unity and Unreal Engine (UE) to obtain full sets of simulated sensors and annotations has become an important direction of development. With the advancement of generation tools, it has become possible to generate multi-task-oriented synthetic datasets. This section summarizes the existing synthetic datasets for multi-task autonomous driving in terms of sensor suites, targeted tasks, domain shifts, etc. We present samples of these datasets in Fig.~\ref{fig:multi-task} and give a comparison in TABLE~\ref{tab2}.

\begin{table*}[htbp]
\begin{center}
\caption{Comparison of Multi-task Synthetic Datasets for Autonomous Driving}
\label{tab2}
\resizebox{\textwidth}{!}{
\begin{tabular}{l*{14}{c}}
\midrule
\textbf{Multi-task}&\textbf{ }&\multicolumn{4}{c}{\textbf{Sensors}}&\multicolumn{8}{c}{\textbf{Tasks}}\\
%&\multicolumn{3}{c}{\textbf{Domain Shifts}}\\
\cmidrule(r){3-6} \cmidrule(r){7-14}
\textbf{} & 
\textbf{\textit{ }}
&\textbf{\textit{Monocular}}& \textbf{\textit{Stereo}}& \textbf{\textit{}}&\textbf{\textit{}}
\\

\textbf{Dataset} & 
\textbf{\textit{Frames}}
&\textbf{\textit{Camera}}& \textbf{\textit{Camera}}& \textbf{\textit{LiDAR}}&\textbf{\textit{Radar}}
&\multicolumn{1}{c}{\textbf{\textit{2D Det.}}}& \textbf{\textit{3D Det.}}& \textbf{\textit{MOT}}
&\textbf{\textit{Seg.}}& \textbf{\textit{Depth}}& \textbf{\textit{Flow}}& \textbf{\textit{Pose}}&\multicolumn{1}{c}
{\textbf{\textit{Pred.}}}\\

%&\textbf{\textit{Angle}}& \textbf{\textit{Lighting}}& \textbf{\textit{Weather}}

\midrule
VKITTI \cite{Gaidon2016VirtualWorldsAP}&17K&\checkmark&-&-&-&\checkmark&-&\checkmark&\checkmark&\checkmark&\checkmark&-&-\\
VIPER \cite{Richter2017PlayingFB}& 250K&\checkmark&-&-&-&\checkmark&\checkmark&\checkmark&\checkmark&-&\checkmark&\checkmark&-\\
PerSIL \cite{Hurl2019PreciseSI}&50K&\checkmark&-&\checkmark&-&\checkmark&\checkmark&\checkmark&\checkmark&-&-&-&-\\
ParallelEye \cite{li2018paralleleye}&40K&\checkmark&-&-&-&\checkmark&-&\checkmark&\checkmark&\checkmark&\checkmark&-&-\\
VKITTI2 \cite{Cabon2020VirtualK2}&17K&\checkmark&\checkmark&-&-&\checkmark&-&\checkmark&\checkmark&\checkmark&\checkmark&-&-\\
SHIFT \cite{Sun2022SHIFTAS}&2.5M&\checkmark&\checkmark&\checkmark&-&\checkmark&\checkmark&\checkmark&\checkmark&\checkmark&\checkmark&\checkmark&\checkmark\\
V2X-Sim \cite{Li2022V2XSimMC}&10K&\checkmark&\checkmark&-&-&\checkmark&\checkmark&-&\checkmark&\checkmark&-&-&-\\
AIODrive \cite{AIODrive}&100K&\checkmark&\checkmark&\checkmark&\checkmark&\checkmark&\checkmark&\checkmark&\checkmark&\checkmark&-&-&\checkmark\\
OPV2V \cite{Xu2021OPV2VAO}&11K&\checkmark&-&\checkmark&-&-&\checkmark&-&\checkmark&-&-&-&\checkmark\\
\bottomrule
%\multicolumn{4}{l}{$^{\mathrm{a}}$Sample of a Table footnote.}
\end{tabular}}
\end{center}
\end{table*}

\subsection{Virtual KITTI}

Virtual KITTI (VKITTI) \cite{Gaidon2016VirtualWorldsAP} was released in 2016 by Gaidon et al., which is one of the earliest works to develop a synthetic dataset for autonomous driving. Based on 5 real data sequences from KITTI, 35 synthetic videos were created using the Unity engine, containing about 17,000 frames of high-resolution images. The sensor suite includes an RGB camera, providing monocular vision images. Each frame in Virtual KITTI contains multi-task annotations of 2D object detection, multi-object tracking, depth estimation, optical flow estimation, and pixel-level semantic and instance segmentation. The domain shifts include different camera angles, lighting conditions, and four weather conditions: clear, cloudy, foggy, and heavy rain.

Virtual KITTI also first proposed a practical definition and a set of experiments to prove effectiveness and transferability across real and synthetic domains.

\subsection{VIPER}

VIsual PERception benchmark (VIPER) \cite{Richter2017PlayingFB} was developed by Richter et al. in 2017. The dataset is annotated using a novel approach in GTA-V to obtain data from the graphics hardware communication of the modern computer game GTA without open source code. VIPER contains 254,064 high-resolution frames and covers a total of 184 kilometers of driving, cycling, and walking under different environmental conditions. Its sensor suite includes cameras with GPS and IMU, providing annotations for 2D/3D target detection, multi-target tracking, optical flow estimation, pixel-level semantic segmentation, dense-level instance segmentation, visual odometry, and relative pose estimation tasks for each frame of images. The domain shifts cover 5 different environmental conditions: daytime (cloudy), sunset, rain, snow, and night, as well as different types of scenes (suburban, downtown, etc.).

{VIPER and other synthetic datasets have already been used with real datasets in training the object instance segmentation method \cite{zhang2020virtual} and the semantic segmentation method \cite{chen2019learning}.} By designing proper domain adaptation methods, models can learn useful information from synthetic data to approach real-world state-of-the-art performance. VIPER, SYNTHIA, and Virtual KITTI are often used as synthetic data sources to interact with real data from KITTI and Cityscapes.

\subsection{PreSIL}

Precise Synthetic Image and LiDAR (PreSIL) \cite{Hurl2019PreciseSI}, published by Hurl et al. in 2019, creates an accurate LiDAR simulator based on GTA-V, which generates LiDAR point cloud images for extracted game data and adds more types of annotation. The PreSIL dataset is equipped with RGB cameras and LiDAR, providing more than 50,000 frames of high-resolution images and LiDAR point cloud data with full-resolution depth information. {However, there is no reflectance value available in GTA-V point clouds, which has a negative impact on point cloud segmentation and detection.} Detailed annotations are oriented toward image depth estimation, semantic segmentation (image), point-by-point segmentation (point cloud), 2D/3D object detection, and multi-object tracking tasks. Domain shifts cover different environmental conditions.

The above synthetic datasets are also used for experiments on bridging the domain gap. Conditional domain normalization (CDN) \cite{su2020adapting} uses existing synthetic Foggy Cityscapes, Virtual KITTI, Synscapes \cite{wrenninge2018synscapes}, SIM10K \cite{johnson2016driving}, and PreSIL as well as real BDD100K, KITTI, and Cityscapes to produce real-to-real and synthetic-to-real adaptation benchmarks.

\subsection{ParallelEye}
{The ParallelEye dataset \cite{li2018paralleleye} was proposed in 2019 by Li et al. It was generated with Unity3D and contains seven sub-datasets with 40,251 frames.} ParallelEye is a purely visual dataset with on-board cameras under different directions, providing annotations of object detection, object traction, depth, optical flow, instance segmentation, and semantic segmentation. The images in ParallelEye also have various environmental conditions of weather (such as sun, rain, fog, etc.) and illumination (sunrise, sunset, etc.) to extend their diversity.

{The researchers conducted domain shift test experiments on the ParallelEye dataset and real datasets such as KITTI, CityScapes, etc. to prove its usefulness. This work is also the foundation of a recent related work ParallelEye pipeline \cite{li2023paralleleye}, which proposed a framework for generating photo-realistic image data for autonomous driving. The researchers also developed ParallelEye-CS for visual intelligence testing later in 2019 \cite{li2019paralleleye}.

\subsection{Virtual KITTI2}

Virtual KITTI2 (VKITTI2) \cite{Cabon2020VirtualK2} was improved by Cabon et al. on the basis of Virtual KITTI in 2020 using an updated version of the Unity engine. Virtual KITTI2 has more realistic image details and features, using the same 5 real data sequences as Virtual KITTI to create a synthetic dataset. Besides the original camera, the sensor suite adds a second-view camera to provide binocular vision, thus providing binocular RGB images and depth images. Virtual KITTI2 includes complete annotations oriented to 2D object detection, multi-object tracking, depth estimation, semantic/instance segmentation, and optical flow. Domain shifts are the same as Virtual KITTI.

Some latter perception works use Virtual KITTI1\&2 as experiment bases, such as 3D detection \cite{ancha2020active}, depth completion \cite{qu2020depth, lagos2022pandepth}, 3D segmentation \cite{jaritz2022cross}, and so on.

\subsection{SHIFT}

SHIFT \cite{Sun2022SHIFTAS} was released by Sun et al. in 2022. As the largest synthetic dataset for autonomous driving, SHIFT was created by Carla Simulator, including 4,800+ sequences captured at 8 different positions and about 2.5M labeled images. The sensor suite contains 11 different sensors: a multi-view RGB camera group of 5 cameras, a stereo RGB camera group, an optical flow sensor, a depth camera, a GNSS sensor, an IMU sensor, and a 128-channel LiDAR sensor that provides visual images and point cloud data. SHIFTS covers annotation for 13 perception tasks: 2D/3D object detection, 2D/3D multi-object tracking, monocular/stereo depth estimation, semantic/instance segmentation, optical flow estimation, point cloud registration, visual mileage estimation, trajectory prediction, and human pose estimation.

SHIFT features for its rich domain shifts include continuous and discrete different scenes, weather, time, traffic, and crowd density, and the changes for each variable are uniformly distributed. It is worth noting that SHIFT provides a benchmark and an experiment basis for domain shift algorithms, and we introduce this part in Section \ref{Evaluation}.

\subsection{V2X-Sim}

Vehicle-to-everything (V2X)-Sim \cite{Li2022V2XSimMC} is the first cooperative perception synthetic dataset for autonomous driving, which was developed by Li et al. based on Carla with a total of 10,000 samples from 100 scenes in 3 Carla towns. For sensors, each vehicle is equipped with 6 RGB cameras in different horizontal angles, a bird's eye view (BEV) perspective camera, a full-view 32-channel LiDAR, a GPS, and an IMU, corresponding with the roadside unit (RSU) sensors of 4 different RGB cameras and LiDAR. Aware that V2X-Sim provides visual images and point cloud data of different horizontal perspectives and BEV perspectives of the vehicle and the roadside, providing an experimental basis for developing and testing collaborative perception algorithms. The dataset provides precise annotations for 3D object detection, depth estimation, and pixel-level semantic segmentation.

Another contribution made by V2X-Sim is the open-source testbed for state-of-the-art collaborative perception algorithms. With sufficient synthetic data and benchmarks for detection, tracking, and segmentation tasks, it stimulates the improvement of V2X algorithms.

\subsection{AIODrive}
All-In-One Drive (AIODrive) \cite{AIODrive} is a multi-task dataset designed especially for high-density long-distance point cloud data by Weng et al. in 2021. The dataset contains 100 video sequences, about 100K labeled images, and point cloud data with a sensor suite of 8 commonly used sensors: RGB camera, stereo camera, depth camera, LiDAR, SPAD-LiDAR, mmwave Radar, IMU, and GPS. In particular, this dataset provides a variety of LiDAR point cloud data with different densities and a special long-distance and high-density type. At the same time, AIODrive is also the first to provide SPAD-LiDAR in the public autonomous driving dataset. Complete annotations for 2D/3D target detection, multi-target tracking, prediction, instance segmentation, and depth estimation are provided. In terms of domain shifts, AIODrive includes rich scene maps, adverse weather, and special scenes with accidents and violations of traffic rules.

\subsection{OPV2V}

The Open Dataset for Perception with V2V communication (OPV2V) \cite{Xu2021OPV2VAO} was proposed by Xu et al. in 2022 and is a large-scale vehicle-to-vehicle collaborative perception dataset. The dataset contains 11,464 frames and 232,913 annotated 3D box diagrams of vehicles and provides a comprehensive benchmark of up to 16 models to evaluate fusion strategies and advanced radar target detection algorithms. The dataset sensors include 4 RGB cameras, a 64-channel LiDAR, a GPS, and an IMU, providing researchers with annotated RGB images, LiDAR point cloud data, and BEV perception images. The OPV2V dataset supports collaborative 3D object detection, BEV semantic segmentation, tracking, and prediction using cameras or LiDAR sensors. At the same time, users can define tasks by adding additional sensors, such as depth estimation, sensor data fusion, etc. The domain shifts cover 70 different scenes, and the map is derived from the geographic information of 8 virtual towns built in Carla and a digital replica of the real Culver City in Los Angeles.

Due to the availability of these datasets, the development of collaborative perception algorithms has been facilitated. In the case of the latency-aware perception system SyncNet \cite{lei2022latency}, it is tested on V2X-Sim to verify its outperformance and robustness. Other work in collaborative perception, such as CoAlign \cite{lu2022robust} and  Where2com \cite{hu2022where2comm} also does experiments on OPV2V, V2X-Sim, and a real-world collaborative dataset DAIR-V2X \cite{yu2022dair}.

\section{Comprehensive Evaluation on Synthetic Datasets} \label{Evaluation}
With the development of synthetic datasets, there is an urgent need for thorough evaluations of the synthetic datasets. These evaluations are critical because synthetic datasets serve as the basis for training and testing in autonomous driving applications. In this section, we provide a comprehensive evaluation for the synthetic datasets. We first present an evaluation framework and summarize common evaluation methods. Then, we give a discussion on the gap test and ways to build the bridge between the synthetic dataset and the real-world dataset. Finally, to address the trustworthiness and safety issues that hinder the deployment of autonomous driving, and to avoid the shortcut learning \cite{geirhos2020shortcut} and long-tail \cite{jain2021autonomy,makansi2021exposing} pitfalls, we emphasize the need to consider these two lessons throughout the data development and system testing process.

\subsection{Evaluation and Comparison}
Synthetic datasets for autonomous driving play a positive role in promoting real-world perception tasks, algorithm optimization, and trustworthiness evaluation. However, there is currently no generally accepted theory for the evaluation criteria of synthetic autonomous driving datasets. Here, we evaluate datasets by considering two categories of elements, which are static elements and interactive elements. See Fig.~\ref{fig:multi-evaluate}.

\begin{figure}[htbp]
\centerline{\includegraphics[width=1\linewidth]{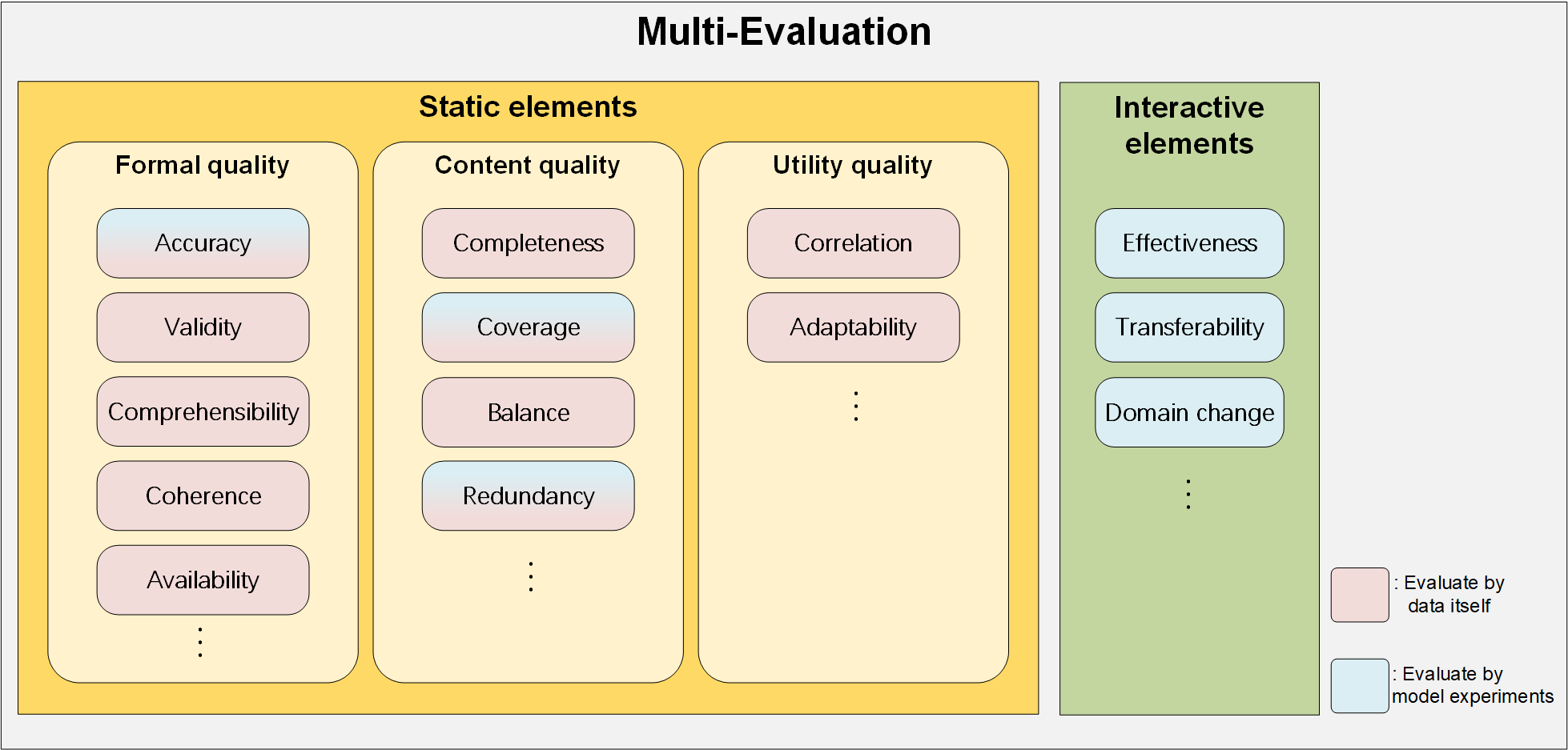}}
\caption{Static elements and interactive elements for multi-evaluating synthetic datasets.}
\label{fig:multi-evaluate}
\end{figure}

\subsubsection{Static elements}
The static elements include the static nature of the data itself and describe the data itself in three aspects: content quality, formal quality, and utility quality.

\textbf{Formal quality} includes the accuracy, validity, comprehensibility, coherence, and availability of the dataset, evaluating the dataset in terms of data expression and form. 
{Accuracy is the fundamental screening of files and data, selecting files that conform to the format to avoid information bias and errors, thereby affecting subsequent work;
Validity primarily considers the degree of alignment between the data itself and its definition, and can evaluate the data from perspectives such as value range and relevance to other parts;
Comprehensibility refers to the ability of the dataset to reflect business logic, and can be evaluated from both subjective and objective perspectives to determine the clarity of fields and values;
Coherence is used to determine whether there is a phenomenon of discontinuity in the data caused by transmission lines, acquisition devices, and other failures;
Availability mainly examines the ease of obtaining the dataset, avoiding the use of unconventional means of acquisition and data that are not based on principles. At the same time, the data should meet the requirements of universality, avoiding extreme and difficult-to-reproduce data.}

\textbf{Content quality} includes the completeness, coverage, balance, and redundancy of the dataset, which mainly describes the specific information contained in the dataset and whether it can meet the requirements of the users.
{Completeness involves the analysis and evaluation of the overall situation of all elements in the scene and can be examined from both static and dynamic aspects of data requirement types.
Coverage, on the other hand, evaluates the completeness of dataset types from a more macro perspective, encompassing both scene coverage and dataset coverage. A good dataset should include various possible scenarios, cover the majority of potential occurrences, and address functionalities like incomplete information, complex environments, and high real-time demands.
Balance revolves around maintaining a proportional balance between evaluation metrics, which is mainly divided into element balance and scene balance. Element balance aims to measure the equilibrium between various elements within the dataset. Scene balance refers to the equilibrium of diverse scenes, including weather conditions, mobile scenarios, extreme situations, etc.
Redundancy examines duplicate parts within the dataset to prevent excessive redundancy from leading to the overfitting of models. This can be assessed from both dynamic and static perspectives, including behavioral and elemental aspects to examine data redundancy. Data redundancy measures the repetition of data, while element redundancy focuses more on redundancy within static elements.}

\textbf{Utility quality} includes correlation and adaptability, which describe demand characteristics and room for development. 
Correlation refers to the degree of correspondence between the concepts and objects described in the dataset and the actual objectives, that is, whether the dataset aligns with the actual objectives and can be effectively applied to real-world goals.
{Adaptability is used to assess the scalability of data and the ability to edit and recombine scenes. Specifically, it can be divided into editability, scalability, and reconstructibility. Editability generally describes the controllable nature of changes in software, data, or computer languages, maintaining certain parts or aspects unchanged while altering others, in order to determine whether usage requirements can be met. Scalability implies the ability to derive or generate new data from existing data through certain means when faced with new task requirements. Reconstructibility refers to the ease with which its internal structure can be adapted, and whether it can meet new requirements by adapting its structure when faced with new testing tasks.}

The quality of static elements can be discussed from two aspects: data acquisition and data processing. The data acquisition level is an assessment of the data itself for the task, including data completeness, label richness, and scene balance \cite{liang2022advances}. Such static elements can usually be evaluated and relatively directly compared from the published information. Whether the type or tag of data can meet the needs of the model training task and whether the background elements, such as weather, surrounding environment, and other scenes, have a balanced distribution, can be figured out from the data description, the sensor suite, domain shifts, and so on.

However, some static elements, such as accuracy and coverage, require manual collation and analysis, but they can also be automatically reflected when applied to specific tasks in machine learning progress. And there are ways to evaluate and improve the static elements score at the data processing level. Common data processing methods include data cleaning and data augmentation. Data cleaning detects harmful sample points (e.g., label errors) and sweeps them out by using confident learning \cite{northcutt2021confident,northcutt2021pervasive} or other frameworks to improve quality. Data augmentation \cite{van2001art,antoniou2017data} has been widely used as a low-cost and effective method to improve the performance and accuracy of machine learning models in a data-constrained environment, which can be achieved by modifying the current sample or fusing multiple samples \cite{liang2021neural}.

\subsubsection{Interactive elements}
The interactive elements describe how the dataset acts on perception and other algorithms or other datasets, such as the performance of training, testing, improving, comparing models, and evaluating the system. On the one hand, the interactive elements are determined by the static elements of the dataset, as the kinds of sensors equipped limit a certain type of support for algorithm experiments. On the other hand, it is also affected by the inherent distribution of the data contained and the gap between the synthetic and real world. How to qualitatively and quantitatively analyze the interactive elements of the synthetic dataset, such as the effectiveness and other effects on the algorithm, generally needs to be evaluated by means of experiments. In terms of the evaluation method, it is mainly to consider the effect of the model, and we discuss several evaluation aspects below.

\textbf{Effectiveness for algorithm training.} As for judging whether the synthetic dataset can effectively train the algorithm model or improve the performance of the algorithm, a rather direct method is to use synthetic datasets to train the model and then test it in a specified real-world environment. However, since there is often a certain data gap between the synthetic data and the real-world data obtained, the training results of the model will often be worse than those of the original real datasets. The Virtual KITTI \cite{Gaidon2016VirtualWorldsAP} dataset was launched with a set of measurement methods proposed, which evaluates the performance of the algorithm by comparing the results of synthetic data training results, real data training results, and pre-training with synthetic data and then fine-tuning with real data results. In comparison, it is demonstrated that synthetic datasets serve as an effective supplement to the real world and can improve the performance of algorithms.

In the SYNTHIA dataset, the balanced gradient contribution (BGC) method proposed by Ros et al. \cite{ros2016training} is used to optimize the algorithm model using synthetic data, and it is proved that synthetic data is an effective simulation and supplement to the real world. Most synthetic datasets, such as VIPER, also use the training performance of benchmark algorithms on their own datasets and other mainstream datasets as a main indicator of measurement and judge the authenticity, challenge, and quality of the dataset by comparing the test results.

\textbf{Transferability of conclusions.} The Virtual KITTI dataset proves the transferability across the synthetic dataset and the real-world results by showing the limited gap in performance on multi-object tracking algorithms between real sequences and cloned sequences. In the SHIFT dataset, the transferability of experimental conclusions from synthetic datasets is evaluated by comparing the trend of the model performance between SHIFT itself and BDD100K under the same discrete domain shifts in multiple perception tasks.

\textbf{Synthetic datasets domain change functionality.} Synthetic datasets can be used to evaluate the robustness and overfitting of the model by testing the degradation of the algorithm under different domain shift subsets. Also, it can be used to test how different data augmentation methods and training strategies reduce the adverse effects of domain shift problems.

The SHIFT dataset tests the performance of multiple perception algorithms using different advanced domain adaptation strategies \cite{Wang2018DeepVD, Li2017DeeperBA, Segu2020BatchNE, Volpi2020ContinualAO, Wang2021TentFT} under discrete and continuous domain shifts, and the expected calibration error (ECE) is estimated by the Softmax \cite{guo2017calibration}, MCDO \cite{gal2016dropout}, and Deep Ensembles \cite{Lakshminarayanan2016SimpleAS} methods. Other methods, such as CDN \cite{su2020adapting} and parallel vision \cite{zhang2020virtual}, use Virtual KITTI, VIPER, etc. to support their results. The above experiments prove that the synthetic dataset provides an experimental platform for domain adaptation algorithm research and emphasize the importance of domain change coverage and fine-grained partitioning for synthetic datasets.

\subsection{Gap Discussion}
The authenticity of synthetic datasets is a topic of much discussion in both academia and industry. These datasets are generated using advanced simulation platforms to produce large amounts of self-labeled data, providing a rapid complement to real-world data collection. However, the challenge of bridging the domain gap between synthetic and real datasets remains. There are two main types of gaps: appearance and content. Appearance gaps refer to pixel- or instance-level differences, such as color, material, texture, and brightness, while content gaps mainly include differences in task label distribution and scene layout \cite{prakash2021self, prabhu2023bridging}. {We also present some methods for filling these gaps based on RGB-D images and point clouds.}

\subsubsection{Appearance gap}We can reduce the appearance gap between synthetic and real-world datasets by using more advanced simulation platforms with high-quality renderers, realistic sensor models, and detailed 3D object models. For example, NVIDIA's DRIVE Sim software utilizes the RTX path-tracing renderer in Omniverse to generate physically accurate sensor data for cameras, radar, LiDAR, and ultrasonic sensors.

To further bridge the gap between synthetic and real-world data, researchers are exploring methods for generating more realistic synthetic datasets, including domain randomization \cite{prakash2021self, tobin2017domain, tremblay2018training, prakash2019structured}, image translation \cite{huang2018multimodal, hoffman2018cycada, chen2019crdoco}, and unsupervised domain adaptation \cite{Tsai2018LearningTA, saito2019strong, zhu2019adapting, hsu2020progressive, yu2022sc}. Specifically, Tremblay et al. \cite{tremblay2018training} proposed a domain randomization technique in which random variations are intentionally introduced into the simulation environment to improve the ability to recognize essential object features. The results show that it is possible to train effective object recognition models using only domain randomized synthetic data, with performance comparable to other processed datasets. Tsai et al. \cite{Tsai2018LearningTA} proposed that images from the source and target domains have high spatial similarity for semantic segmentation. Based on this idea, they present a domain adaptation algorithm that uses adversarial learning in the output space to predict the structured output at the pixel level with spatial and local information. The simulation results show that the method can effectively adapt to changes in the domain, such as synthesis to real data and between different cities.

\subsubsection{Content gap} Simulators also provide solutions, such as the DRIVE Sim simulation software, which includes various 3D object models and domain randomization functions to organize different data generation scenarios. However, we still need to improve the domain adaptation of the generated datasets and ensure that they have good generalization capabilities even in real-world image data. Commonly used techniques to fill the content gap include self-training with pseudo-labels \cite{yu2022sc, tan2020class, li2019bidirectional} and automatic learning to generate scenes \cite{Kar2019MetaSimLT, devaranjan2020meta, tan2021scenegen}. In particular, Yu et al. \cite{yu2022sc} performed self-training with optimized pseudo-labels and error-corrected training procedures. The training process consists of pseudo-label initialization, self-training, and iteration with better labels to achieve annotation of the dataset with high-quality pseudo-labels in the target domain. The pseudo-labels are distributed in the target domain with approximate content, thus helping to reduce the content gap. Kar et al. \cite{Kar2019MetaSimLT} proposed the Meta-Sim approach to obtain simulated data comparable to real data. The approach uses a neural network to parameterize the dataset generator, learn how to modify the properties of the scene graph (probabilistic scene grammar), and reduce the gap between the distribution of the generated data and the real data, resulting in improved image quality.

\subsubsection{Intermediary}

In general, the data in synthetic datasets and real datasets are distributed in two domains with large differences, respectively. Although the domain gap can be narrowed by the specific means described above, the model trained using the synthetic datasets still needs to be fine-tuned on real datasets to learn the features of both domains simultaneously and apply them to real scenarios. However, this pre-training and fine-tuning learning model also has disadvantages, such as wasting computational resources and producing less-than-optimal learning results. If a high-fidelity virtual 3D reconstruction or rendering can be performed based on the actual acquired data, virtual data with a very similar distribution to the real data can be easily generated. The models trained on such data can be directly applied to real scenes without the procedure of pre-training with synthetic data and fine-tuning with real data.

The sensors used in autopilot datasets are multimodal, so the collected data are mainly based on pictures taken by RGB-D cameras and point clouds captured by Radars and LiDARs. In the process of generating synthetic datasets, the information provided by the images can be used to generate new perspective images, but the disadvantage is that the 2D images do not provide comprehensive 3D information and are easily affected by occlusions. Although the point clouds can provide intuitive 3D structure information, the shortcoming is the lack of semantic information and the very sparse and inaccurate point clouds generated by outdoor scenes.

\textbf{RGB-D Images.} For multi-angle images taken by in-car cameras or roadside cameras, neural radiance field (NeRF) and its variants \cite{mildenhall2021nerf, zhang2020nerf++, muller2022instant, tancik2022block, chen2021mvsnerf} can perform data generation by re-rendering different angles of the scene using an implicit 3D representation. NeRF is a recently proposed method for 3D scene reconstruction and new view synthesis. The main idea of NeRF is to represent the scene as a continuous 3D function that can evaluate any point in space to obtain the corresponding color and opacity. The function is modeled by a neural network that is trained on a set of input images and corresponding camera poses. NeRF and its variants represent a significant advance in 3D scene reconstruction and novel view synthesis using deep learning. The ability to represent scenes as continuous functions and capture fine details as well as lighting and texture changes has made NeRF a novel technology for a wide range of applications in virtual reality, gaming, robotics, etc.

\textbf{Point Clouds.} Since the point cloud data obtained from the scanned outdoor scenes by vehicle-mounted LiDAR is at least one order of magnitude sparser than the indoor scenes, semantic scene complementation becomes very important to recover the 3D scene representation. For the first time, SSCNet \cite{song2017semantic} combines 3D Shape Completion and Semantic Scene Labeling together and demonstrates experimentally that these two different tasks can inform each other and that an end-to-end algorithm that accomplishes both tasks simultaneously works better than either scene reconstruction or semantic segmentation alone. However, the mainstream input form of the series of 3D CNN based work \cite{li2019rgbd, li2019depth, li2020anisotropic} is 2D images such as range images or RGB-D images, while S3CNet \cite{cheng2021s3cnet} targets sparse point clouds in large scenes for semantic scene complementation, and both types of methods play an important role in refining the collection of synthetic datasets.

In addition, some migration synthesis can be performed using desktop datasets \cite{firman2016structured, xu2022scene} or 3D shape datasets \cite{chang2015shapenet} to enrich the autopilot synthesis of outdoor datasets. For example, using methods such as truncated signed distance function (TSDF) \cite{curless1996volumetric} for general objects and methods such as skinned multi-person linear (SMPL) model \cite{loper2015smpl} or implicit field \cite{saito2019pifu} for human bodies, 3D representations of desktops, indoor objects, or pedestrians can be reconstructed and migrated to outdoor scenes. With ShapeNet \cite{chang2015shapenet}, objects that do not normally appear in autopilot scenes can also be migrated to autopilot synthetic datasets to solve the long-tail distribution problem of natural data and enrich the edge data distribution. This not only enriches the range of data encompassed by autopilot datasets but also improves the flexibility of their generation.

\subsection{Trustworthiness and Safety}
Trustworthiness and safety have always been the top priorities of autonomous driving algorithms for researchers. Due to the special application scenarios of autonomous driving, a tiny mistake will pose a serious threat to personal safety and property. In the National Artificial Intelligence Research and Development Strategic Plan (US) \cite{national2019national}, the effective evaluation method of autonomous driving algorithms is also listed as one of the most important issues.

An important way to evaluate the trustworthiness of AI is to conduct reasonable tests and experiments based on data. In the face of increasingly advanced artificial intelligence algorithms, more researchers have realized that the quality of the data greatly affects the generalizability and reliability of the model. The fake correlations and biases learned in datasets and their labels seriously decrease credibility \cite{xu2022safebench}. Therefore, it is of great value to focus on data design and data generation pathways so as to reliably monitor and evaluate the generation of models. In terms of trustworthiness, good coverage of distribution and fine-grained division of labels are of great significance for subsequent testing, for which the establishment of a sustainable iterative and community-built dataset platform is an important means to meet the above requirements.

As to the evaluation method of trustworthiness, the idea of ablation experiments can be used to evaluate the quality of data by removing different subsets in model training \cite{liang2022advances} and observing the performance changes of AI, such as calculating the Shapley value of the dataset for fair valuation \cite{ghorbani2019data, kwon2021efficient,jia2019towards,koh2017understanding}. The cleaning of the dataset can also be done by predicting the uncertainty of the data \cite{northcutt2021confident} so as to improve the credibility of the training model. In addition, the annotation of synthetic autonomous driving data comes from the automatic generation of the scene model. Compared with manual annotation, it not only reduces the cost but also reduces the bias error of human work, and with the method of aggregating labels to reduce noise, trustworthiness is thereby further improved. {SE method is also proposed by Li et al. to achieve trust in AI, including intelligence and index (I\&I) to ensure the quality and trust, calibration and certification (C\&C) to ensure the result quality and some performance, such as security, and verification and validation (V\&V) to ensure the right operational flow and output \cite{li2022features}. Related work of SEEFMM framework with six levels for foundation trust models is proposed in 2023 \cite{li2022novel}.}

For the safety evaluation of autonomous driving algorithms, using naturally collected datasets faces a serious long-tail problem in important scenarios. According to the California Department of Motor Vehicles, critical scenarios for safe driving occur once every 30,000 miles driven on average \cite{california}. Such a low frequency makes the safety test often require huge economic and time costs, especially since such a long-tailed distribution requires the method of assessment to converge according to probability to ensure the confidence of the results. Some important methods, such as importance sampling (IS) \cite{zhao2016accelerated,o2018scalable} and oversampling \cite{wallace2011class}, deal with the long-tail distribution with optimizing sampling and distribution density estimation techniques. Other methods balance rare labels in the training process by adjusting cross-entropy (CE) \cite{arief2021deep}, logit functions \cite{menon2020long} or normalized weights \cite{kim2020adjusting}. However, the above methods often hit a plateau in the face of more complex distribution problems and struggle with underestimating security risks. Other direct ways to ensure safety in daily driving are to optimize coverage of real datasets by expanding interesting scenarios \cite{mao2021one} or scoring scenes to bias data selection from long-tail distribution \cite{wilson2023argoverse}, though expensive experiments are needed. These ideas can also be adopted by synthesizing data and avoiding drawbacks.

Therefore, another resolution is to especially design important scene data for safety testing with generalization capabilities, such as the prior work of using adversarial generation \cite{ding2021multimodal,zhang2022adversarial,feng2021intelligent} or knowledge \cite{ding2021semantically,bagschik2018ontology} to synthesize task-focused safety testing datasets. At present, there are also platforms that provide experimental conditions for this type of test. For example, the SafeBench platform provides 2,352 generated safety-critical scenarios (such as benign scenarios of straight obstacles and lane changes) in safety tests, with the evaluation results based on ten metrics such as collision rate, frequency of running red lights, average path completion rate, etc. \cite{xu2022safebench}.

\section{Experiments} \label{Experiments}
{In this section, we conduct a series of evaluation experiments on dataset samples. We show the evaluation results over examples of multi-evaluation, safety and trustworthiness, and then indicate the possible supplementary generation aspects to improve the quality of datasets.}

\subsection{Static Element Evaluation}
{To demonstrate the effectiveness of the multi-evaluation framework, we first select two synthetic autonomous driving datasets for static element evaluation, namely the Virtual KITTI dataset and the SHIFT dataset. Using our evaluation approach, we employ the Analytic Network Process (ANP) \cite{saaty2006decision,saaty2013analytic} to analyze the relationships among the indicators of form quality, content quality, and utility quality, forming an indicator relationship network, thereby calculating the weights. We assign a maximum score of 5.00 to each criterion and determine the objective scores based on the definitions and calculation equations provided for the aforementioned indicators. By applying the weights generated using the ANP method, we can compute an all-encompassing score that assesses various autonomous driving datasets comprehensively. And the evaluation results and indicator weights are shown in TABLE~\ref{tab:experiment_result}.}

\begin{table*}[htbp]
\begin{center}
\caption{Static Element Evaluation on Virtual KITTI and SHIFT}
\label{tab:experiment_result}
\begin{threeparttable}
\begin{tabular}{l*{13}{c}}
\toprule
\textbf{Autonomous}&\multicolumn{5}{c}{\textbf{Form}}&\multicolumn{4}{c}{\textbf{Content}}
&\multicolumn{2}{c}{\textbf{Utility}}&\multicolumn{2}{c}{\textbf{Overall}}\\

\textbf{Dataset} & 
\textbf{Acc}
&\textbf{Val}& \textbf{Com}& \textbf{Con}&\textbf{Ava}&\textbf{E.C.}&\textbf{Cov}&\textbf{Bal}&\textbf{Red}&\textbf{Cor}&\textbf{Ada}&\textbf{Value}\\

\midrule
Virtual KITTI & 5.00 & 5.00 & 5.00 & 5.00 & 4.80 & 3.84 & 4.52 & 4.24 & 4.78 & 4.75 & 4.50 & 4.73\\
SHIFT & 5.00 & 5.00 & 5.00 & 5.00 & 4.60 & 4.67 & 4.80 & 4.58 & 4.71 & 4.90 & 4.60 & 4.84 \\
\midrule
Weight & 0.061 & 0.040 & 0.176 & 0.064 & 0.034 & 0.039 & 0.046 & 0.126 & 0.121 & 0.148 & 0.152 & 1.000\\
\bottomrule
%\multicolumn{4}{l}{$^{\mathrm{a}}$Sample of a Table footnote.}
\end{tabular}
\begin{tablenotes}

\item[*] Notion of indicators: Accuracy (Acc), Validity (Val), Comprehensibility (Com), Continuity (Con), Availability (Ava), Element completeness (E.C.), Coverage (Cov), Balance (Bal), Redundancy (Red), Correlation (Cor), Adaptability (Ada). 
\end{tablenotes}
\end{threeparttable}
\end{center}
\end{table*}

In terms of formal quality, for accuracy, effectiveness, and continuity, we choose the ratio of the correct number in the dataset to the total number as its score. For comprehensibility, we use quality assessments of individual images, such as PSNR and SSIM \cite{hore2010image}, to calculate it, and obtain availability by calculating its time cost and space cost, using the Analytic Hierarchy Process \cite{saaty1988analytic,saaty2008decision} to derive its score. In terms of content quality, we manually calculate the distribution of various elements in relation to the scene to determine completeness and coverage scores, and synthesize the results to obtain a coherence score. In terms of redundancy, we calculate the percentages of time redundancy and element redundancy, and take a weighted average to obtain the corresponding quality score. In terms of utility quality, correlation shows whether the dataset can meet practical requirements and be effectively and fully utilized for the actual objectives which can be determined by the proportion of effective datasets among all datasets based on specific perception tasks. In terms of adaptability, performance degradation when using the current dataset to train a model for application to other datasets is compared to serve as an evaluation criterion. For the Analytic Network Process, relationships of importance among different weights and the influence of elements are set based on user requirements, and Super Decision (SD) software \cite{adams2003super} is used to obtain the weight of each indicator, thus obtaining the overall quality of the dataset.

{According to the table, we can draw some conclusions. Overall, the quality of the SHIFT dataset is higher than that of the Virtual KITTI dataset. First, both of the datasets are excellent in form quality, for autonomous driving synthetic datasets, it is not common to encounter formal errors, and they are generally more cost-effective compared to real-world datasets. However, compared with Virtual KITTI, SHIFT dataset is much larger and may need to change the weather according to the need of the users, so the availability is lower than that of Virtual KITTI. In terms of the content quality, the main drawback of the Virtual KITTI dataset is that it includes only vehicles and lacks other traffic participants such as pedestrians and cyclists, and it does not include nighttime data. These scenarios are frequently encountered in driving. In comparison, the SHIFT dataset shows superior performance. However, when it comes to utility quality, the differences between synthetic and real-world datasets are inevitable. Synthetic datasets tend to be less scalable and less relevant. To apply synthetic datasets in real-world scenarios, it is necessary to compensate for these gaps to ensure the usability of the model.}

\subsection{Interactive Element Evaluation and Safety Test}
{As for the interactive elements, we also perform a domain shift test on Virtual KITTI to estimate its transferability and safety under different conditions. We train a 2D vehicle detection model YOLOv5s on videos 0001, 0002, and 0006, and test it on Virtual KITTI 0020, a random subset of 3,769 images from KITTI and about 10,000 images from the BDD100K validation set. To explore the domain shift performance in more detail, we divide the images in the BDD100K validation set by time of day and weather (excluding conditions with images fewer than 50 images). The results are shown in Fig.~\ref{fig:bddmap}. We also calculate the relative robustness performance (R) defined as $R = \frac{Shift\ Performance}{Original\ Performance}$ \cite{yu2023benchmarking} to show the impact of different conditions and the R matrix is shown in TABLE~\ref{tab:bdd100k}. Note that the model performs best in the original Virtual KITTI test set. As Virtual KITTI imitates the KITTI dataset, it also has a relatively good performance on KITTI. For the BDD100K dataset transferability experiment, the performance drops due to the domain gap. The worst three performances are marked in red, and obviously, these are all night scenes in BDD100K. As mentioned above, Virtual KITTI does not have data for night scenes, so the result is reasonable. It also indicates that in order to further improve the quality of Virtual KITTI and make it safer for a wider range of conditions, supplementing night scenes could be of good effectiveness.}

\begin{figure}[htbp]
\centerline{\includegraphics[width=1\linewidth]{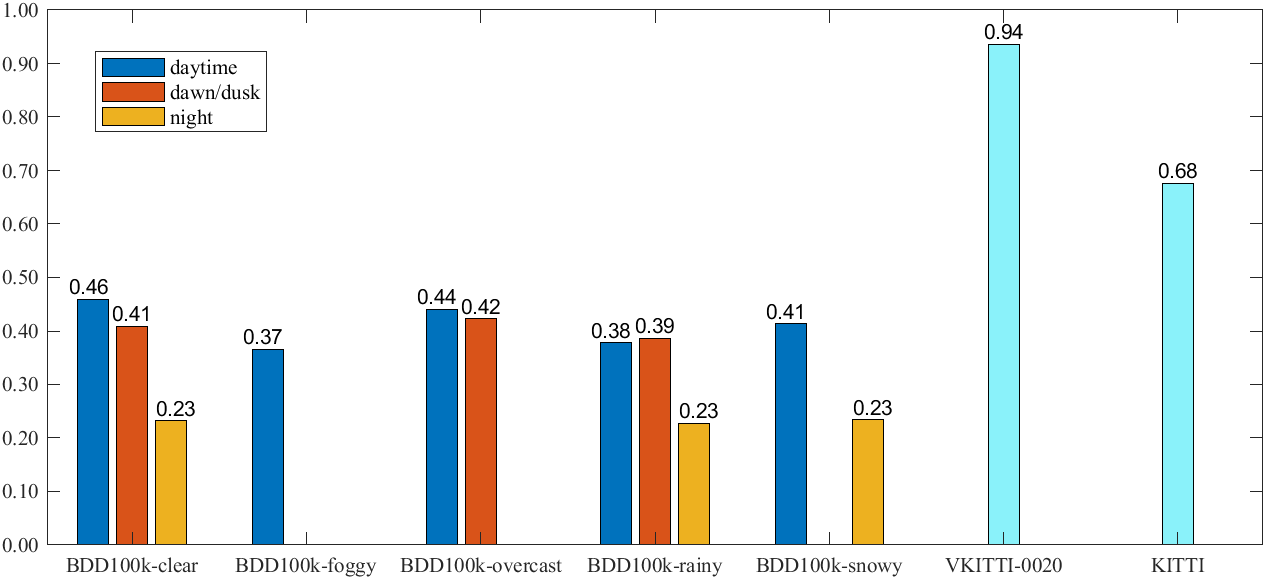}}
\caption{Results of domain shift on KITTI and different situations in BDD100K (mAP50). We trained a YOLOv5s car detection model on Virtual KITTI training set and tested its performance on KITTI and different time and weather data splits of the BDD100K validation set.  }
\label{fig:bddmap}
\end{figure}

\begin{table}[htbp]
\begin{center}
\caption{Relative Robustness of Virtual KITTI Model under Different Conditions in BDD100K.}
\label{tab:bdd100k}
\begin{threeparttable}
\begin{tabular}{l*{4}{c}}
\toprule
\diagbox{Weather}{Time of day} & \textbf{Daytime} & \textbf{Dawn/Dusk} & \textbf{Night} \\
\midrule
\textbf{Clear} & 0.491 & 0.437 & \textcolor{red}{0.249}\\
\textbf{Foggy} & 0.392 & - & -\\
\textbf{Overcast} & 0.471 & 0.452 & -\\
\textbf{Rainy} & 0.404 & 0.413 & \textcolor{red}{0.244}\\
\textbf{Snowy} & 0.442 & - & \textcolor{red}{0.250}\\

\bottomrule

\end{tabular}
\end{threeparttable}
\end{center}
\end{table}

\subsection{Trustworthiness Test }
{We apply a data-clean algorithm as an example of a trustworthiness test, because although the synthetic dataset has received automatic labels based on object positions, sometimes there may exist obstructions and incompleteness (such as a person behind a truck thus causing a labeling problem). Specifically, we use Cleanlab \cite{northcutt2021confident} to evaluate the quality of labels in Virtual KITTI. This method can use both pre-trained models on other datasets like KITTI and self-trained models for cross-validation. From such methods, we can go back to remove inappropriate annotations to further improve the trustworthiness of synthetic datasets.}

{See TABLE~\ref{tab:clean-data}. We use Virtual KITTI and KITTI as experimental sources. We first train two models from Virtual KITTI and KITTI separately. Then we use these two models to evaluate the label equality of the two test subsets. Since Cleanlab is designed for image classification tasks, we use origin labels to divide the images into object patches and then apply the model to infer class probabilities to conduct the evaluation. As shown in TABLE~\ref{tab:clean-data}, the labels in Virtual KITTI-v0020 are carefully selected (in the DontCare class) and clean, which proves the advantage of data quality in synthetic datasets. To further explore the usability of this method, we randomly change 100 labels in Virtual KITTI-v0020 (referred to as Corrupted-v0020) into wrong labels, and Cleanlab reports 84 of them in 459 label issues with KITTI trained model and 96 out 96 with Virtual KITTI model. For the KITTI dataset, we use the test set for the KITTI trained model to do the clean work and report 709 label issues in 3,769 images. In these reported issues, some are blurred and heavily obstructed (might be labeled as DontCare), some have label position errors, some are wrongly labeled and many of them are also misreported. See Fig.~\ref{fig:cleandata}. Therefore, such trustworthiness tests and manual verification can improve the trustworthiness of synthetic datasets.}

\begin{table*}[htbp]
\begin{center}
\caption{Comparison of Data-Clean Process Experiment}
\label{tab:clean-data}
\begin{threeparttable}
\begin{tabular}{l*{6}{c}}
\toprule
\textbf{Dataset}&\textbf{Model}&\textbf{Total Label}&\textbf{True Label Issue}&\textbf{Report Label Issue}&\textbf{Label Health Score}\\

\midrule
Virtual KITTI-v0020 & KITTI & 31,775 & - & 0 & 1.00\\
Virtual KITTI-v0020 & Virtual KITTI & 31,775 & - & 0 & 1.00\\
Corrupted-v0020 & KITTI & 31,775 & 100 & 459 & 0.99\\
Corrupted-v0020 & Virtual KITTI & 31,775 & 100 & 96 & 1.00\\
KITTI-test & KITTI & 17,558 & - & 709 & 0.96\\

\bottomrule

\end{tabular}
\end{threeparttable}
\end{center}
\end{table*}

\begin{figure}[htbp]
\centerline{\includegraphics[width=1\linewidth]{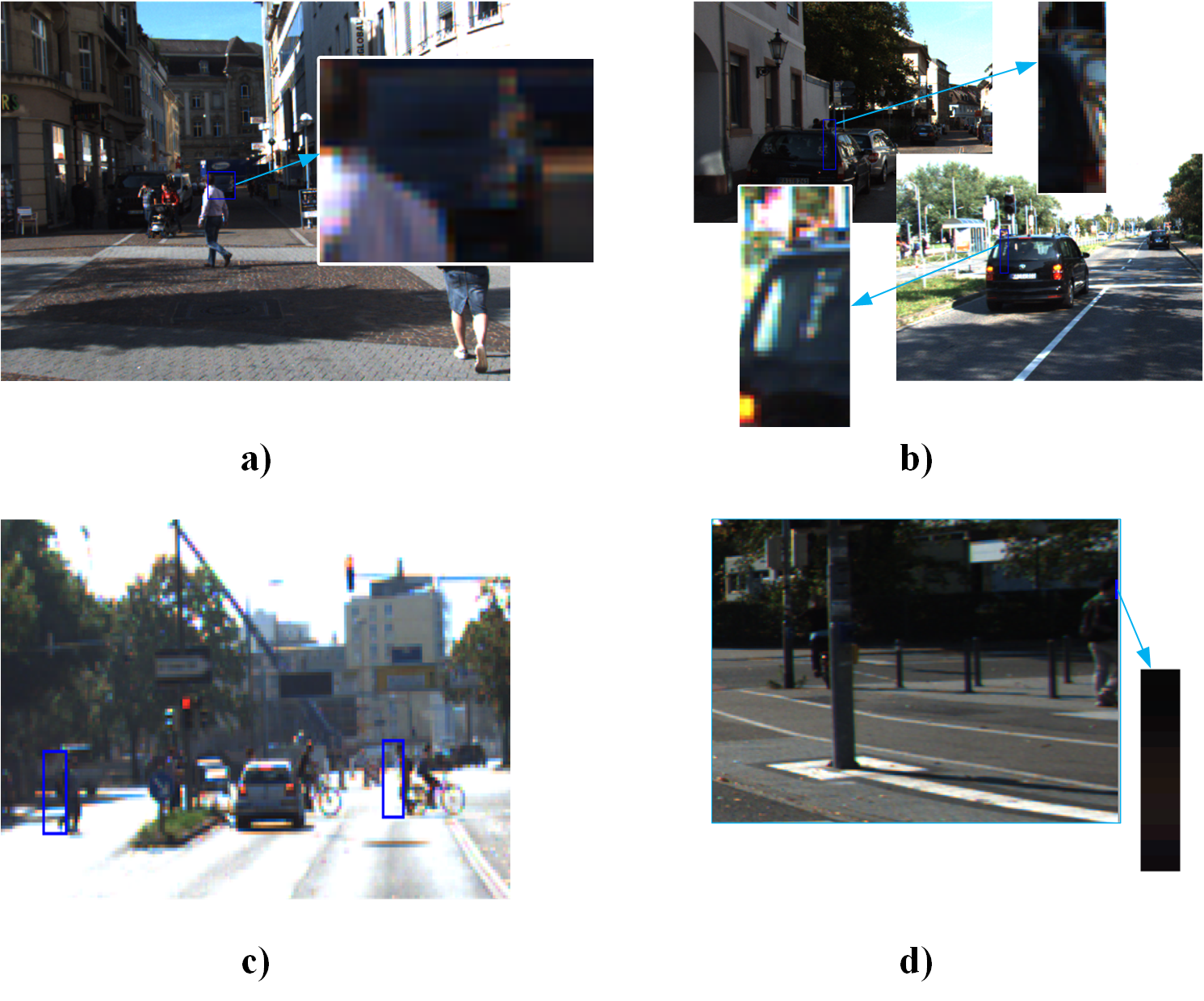}}
\caption{Samples of reported label issues in KITTI. a) Heavily Blurred, b) Serious occlusion, c) Position Error, d) Wrong Label (Labeled as pedestrian).}
\label{fig:cleandata}
\end{figure}

\section{Disscusions and Future directions} \label{Future}
{Synthetic datasets have shown great promise in the development process, but they are not without limitations. A major challenge is to adequately represent the complexity of the real world. For example, synthetic data may struggle to capture the nuances of changing weather conditions or highly dynamic traffic scenarios. Researchers should focus on increasing the realism and diversity of synthetic datasets. This can be achieved through improved data generation techniques, more sophisticated simulations, and the incorporation of real-world data for calibration. On the other hand, the majority of synthetic datasets in the autonomous driving domain focus on environmental perception tasks, leaving a lack of datasets designed for driver behavior analysis and decision learning. More synthetic datasets should be designed to address these issues.}

In the ever-advancing field of autonomous driving research, the availability of high-quality datasets is critical. To comprehensively address this need and overcome the limitations of synthetic datasets, we propose a systematic process for synthetic dataset generation. As shown in Fig.~\ref{fig:conclusion}, the process aims to produce trustworthy synthetic datasets that can be effectively evaluated and continuously improved to meet to the specific requirements of autonomous driving research and development. This will guide the direction of synthetic datasets.

\begin{figure*}[htbp]
\centerline{\includegraphics[width=0.8\linewidth]{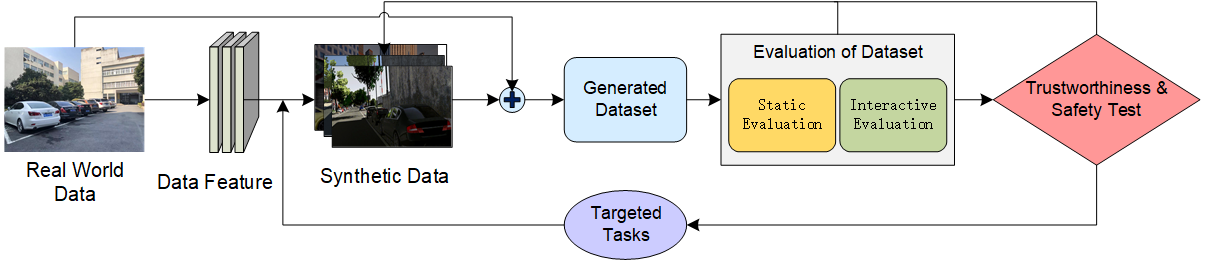}}
\caption{A systematic process of generating trustworthy data for autonomous driving.}
\label{fig:conclusion}
\end{figure*}

{The innovation of our systematic process is the inclusion of feedback loops. We believe that the dataset generation is not a one-shot work, but the evaluation and regeneration work should be combined in the whole development process. After the generation step, the synthetic datasets are used for comprehensive evaluations, including the multi-evaluation, trustworthiness, and safety tests mentioned in Section \ref{Experiments} to evaluate their quality, and then for training and evaluating autonomous driving algorithms on targeted tasks. In addition, the results of evaluation and model use are fed back into the dataset generation step, initiating an iterative improvement process. The evaluation results, such as the experimental data from on-road autonomous vehicle tests, inform adjustments and enhancements to the datasets, addressing any shortcomings or discrepancies identified during the evaluation phase. This iterative approach ensures that the synthetic datasets are continuously consummated to better reflect the complexity of real-world driving scenarios and contribute to the development of perception algorithms.}

Additionally, it is necessary for future researchers to explore how to close the gap between synthetic and real-world datasets and to evaluate the usefulness of generated data. As we emphasized trustworthiness and safety in marketing the product of autonomous driving, data development should be applied to the entire process of testing autonomous vehicles to help with iteration. How to expand the coverage to corner cases and solve the long-tail problems are also of great value.

\section{Conclusion} \label{Conclusion}
This paper provides an overview of synthetic datasets for autonomous vehicles. It first outlines the process of generating these datasets and categorizes them as either single-task or multi-task. Then we propose a practical way to evaluate datasets in two aspects and discuss the evaluation of usefulness shown in model training, bridging the gap between synthetic and real-world datasets. The challenge of evaluating these synthetic datasets is then explored, including strategies to bridge the gap between synthetic and real-world data. We conclude the study by emphasizing the importance of synthetic datasets for trustworthiness and safety in autonomous driving and introducing several possible promising directions for synthetic data generation.

% Can use something like this to put references on a page
% by themselves when using endfloat and the captionsoff option.
\ifCLASSOPTIONcaptionsoff
  \newpage
\fi

% trigger a \newpage just before the given reference
% number - used to balance the columns on the last page
% adjust value as needed - may need to be readjusted if
% the document is modified later
%\IEEEtriggeratref{8}
% The "triggered" command can be changed if desired:
%\IEEEtriggercmd{\enlargethispage{-5in}}

% references section
\bibliographystyle{IEEEtran} %声明选择的格式
\bibliography{references} % bib文件名，需要放在同一个文件夹下，否则要在filename前说明路径

% Generated by IEEEtran.bst, version: 1.14 (2015/08/26)
\begin{thebibliography}{100}
\providecommand{\url}[1]{#1}
\csname url@samestyle\endcsname
\providecommand{\newblock}{\relax}
\providecommand{\bibinfo}[2]{#2}
\providecommand{\BIBentrySTDinterwordspacing}{\spaceskip=0pt\relax}
\providecommand{\BIBentryALTinterwordstretchfactor}{4}
\providecommand{\BIBentryALTinterwordspacing}{\spaceskip=\fontdimen2\font plus
\BIBentryALTinterwordstretchfactor\fontdimen3\font minus \fontdimen4\font\relax}
\providecommand{\BIBforeignlanguage}[2]{{%
\expandafter\ifx\csname l@#1\endcsname\relax
\typeout{** WARNING: IEEEtran.bst: No hyphenation pattern has been}%
\typeout{** loaded for the language `#1'. Using the pattern for}%
\typeout{** the default language instead.}%
\else
\language=\csname l@#1\endcsname
\fi
#2}}
\providecommand{\BIBdecl}{\relax}
\BIBdecl

\bibitem{doi:10.1177/0278364913491297}
A.~Geiger, P.~Lenz, C.~Stiller, and R.~Urtasun, ``Vision meets robotics: The kitti dataset,'' \emph{The International Journal of Robotics Research}, vol.~32, no.~11, pp. 1231--1237, 2013.

\bibitem{sun2020scalability}
P.~Sun, H.~Kretzschmar, X.~Dotiwalla, A.~Chouard, V.~Patnaik, P.~Tsui, J.~Guo, Y.~Zhou, Y.~Chai, B.~Caine \emph{et~al.}, ``Scalability in perception for autonomous driving: Waymo open dataset,'' in \emph{Proceedings of the IEEE/CVF conference on computer vision and pattern recognition}, 2020, pp. 2446--2454.

\bibitem{Yu2018BDD100KAD}
F.~Yu, H.~Chen, X.~Wang, W.~Xian, Y.~Chen, F.~Liu, V.~Madhavan, and T.~Darrell, ``Bdd100k: A diverse driving dataset for heterogeneous multitask learning,'' in \emph{Proceedings of the IEEE/CVF conference on computer vision and pattern recognition}, 2020, pp. 2636--2645.

\bibitem{he2022synthetic}
R.~He, S.~Sun, X.~Yu, C.~Xue, W.~Zhang, P.~Torr, S.~Bai, and X.~QI, ``Is synthetic data from generative models ready for image recognition?'' in \emph{The Eleventh International Conference on Learning Representations}, 2022.

\bibitem{paulin2023review}
G.~Paulin and M.~Ivasic-Kos, ``Review and analysis of synthetic dataset generation methods and techniques for application in computer vision,'' \emph{Artificial Intelligence Review}, pp. 1--45, 2023.

\bibitem{nikolenko2021synthetic}
S.~I. Nikolenko, \emph{Synthetic data for deep learning}.\hskip 1em plus 0.5em minus 0.4em\relax Springer, 2021, vol. 174.

\bibitem{huffman1971impossible}
D.~A. Huffman, ``Impossible objects as nonsense sentences,'' \emph{Machine intelligence}, vol.~6, pp. 295--323, 1971.

\bibitem{lucas1981iterative}
B.~D. Lucas and T.~Kanade, ``An iterative image registration technique with an application to stereo vision,'' in \emph{IJCAI'81: 7th international joint conference on Artificial intelligence}, vol.~2, 1981, pp. 674--679.

\bibitem{gers2001lstm}
F.~A. Gers and E.~Schmidhuber, ``Lstm recurrent networks learn simple context-free and context-sensitive languages,'' \emph{IEEE transactions on neural networks}, vol.~12, no.~6, pp. 1333--1340, 2001.

\bibitem{schneider2016novo}
P.~Schneider and G.~Schneider, ``De novo design at the edge of chaos: Miniperspective,'' \emph{Journal of medicinal chemistry}, vol.~59, no.~9, pp. 4077--4086, 2016.

\bibitem{fadaee2017data}
M.~Fadaee, A.~Bisazza, and C.~Monz, ``Data augmentation for low-resource neural machine translation,'' \emph{arXiv preprint arXiv:1705.00440}, 2017.

\bibitem{jpt-iv10}
J.-P. Tarel, N.~Hautiere, A.~Cord, D.~Gruyer, and H.~Halmaoui, ``Improved visibility of road scene images under heterogeneous fog,'' in \emph{2010 IEEE intelligent vehicles symposium}.\hskip 1em plus 0.5em minus 0.4em\relax IEEE, 2010, pp. 478--485.

\bibitem{jpt-itsm12}
J.-P. Tarel, N.~Hautiere, L.~Caraffa, A.~Cord, H.~Halmaoui, and D.~Gruyer, ``Vision enhancement in homogeneous and heterogeneous fog,'' \emph{IEEE Intelligent Transportation Systems Magazine}, vol.~4, no.~2, pp. 6--20, 2012.

\bibitem{Butler2012ANO}
D.~J. Butler, J.~Wulff, G.~B. Stanley, and M.~J. Black, ``A naturalistic open source movie for optical flow evaluation,'' in \emph{Computer Vision--ECCV 2012: 12th European Conference on Computer Vision, Florence, Italy, October 7-13, 2012, Proceedings, Part VI 12}.\hskip 1em plus 0.5em minus 0.4em\relax Springer, 2012, pp. 611--625.

\bibitem{CC3}
``Creative commons attribution 3.0 license,'' \url{https://creativecommons.org/licenses/by/3.0/legalcode}.

\bibitem{mayer2016large}
N.~Mayer, E.~Ilg, P.~Hausser, P.~Fischer, D.~Cremers, A.~Dosovitskiy, and T.~Brox, ``A large dataset to train convolutional networks for disparity, optical flow, and scene flow estimation,'' in \emph{Proceedings of the IEEE conference on computer vision and pattern recognition}, 2016, pp. 4040--4048.

\bibitem{7301289}
M.~Savva, A.~X. Chang, and P.~Hanrahan, ``Semantically-enriched 3d models for common-sense knowledge,'' in \emph{Proceedings of the IEEE Conference on Computer Vision and Pattern Recognition Workshops}, 2015, pp. 24--31.

\bibitem{101007}
S.~R. Richter, V.~Vineet, S.~Roth, and V.~Koltun, ``Playing for data: Ground truth from computer games,'' in \emph{Computer Vision--ECCV 2016: 14th European Conference, Amsterdam, The Netherlands, October 11-14, 2016, Proceedings, Part II 14}.\hskip 1em plus 0.5em minus 0.4em\relax Springer, 2016, pp. 102--118.

\bibitem{Gaidon2016VirtualWorldsAP}
A.~Gaidon, Q.~Wang, Y.~Cabon, and E.~Vig, ``Virtual worlds as proxy for multi-object tracking analysis,'' in \emph{Proceedings of the IEEE conference on computer vision and pattern recognition}, 2016, pp. 4340--4349.

\bibitem{Isola2016ImagetoImageTW}
P.~Isola, J.-Y. Zhu, T.~Zhou, and A.~A. Efros, ``Image-to-image translation with conditional adversarial networks,'' in \emph{Proceedings of the IEEE conference on computer vision and pattern recognition}, 2017, pp. 5967--5976.

\bibitem{Wang2017HighResolutionIS}
T.-C. Wang, M.-Y. Liu, J.-Y. Zhu, A.~Tao, J.~Kautz, and B.~Catanzaro, ``High-resolution image synthesis and semantic manipulation with conditional gans,'' in \emph{Proceedings of the IEEE conference on computer vision and pattern recognition}, 2018, pp. 8798--8807.

\bibitem{Chen2017PhotographicIS}
Q.~Chen and V.~Koltun, ``Photographic image synthesis with cascaded refinement networks,'' in \emph{Proceedings of the IEEE international conference on computer vision}, 2017, pp. 1520--1529.

\bibitem{Qi2018SemiParametricIS}
X.~Qi, Q.~Chen, J.~Jia, and V.~Koltun, ``Semi-parametric image synthesis,'' in \emph{Proceedings of the IEEE Conference on Computer Vision and Pattern Recognition}, 2018, pp. 8808--8816.

\bibitem{Park2019SemanticIS}
T.~Park, M.-Y. Liu, T.-C. Wang, and J.-Y. Zhu, ``Semantic image synthesis with spatially-adaptive normalization,'' in \emph{Proceedings of the IEEE/CVF conference on computer vision and pattern recognition}, 2019, pp. 2337--2346.

\bibitem{Tang2019MultiChannelAS}
H.~Tang, D.~Xu, N.~Sebe, Y.~Wang, J.~J. Corso, and Y.~Yan, ``Multi-channel attention selection gan with cascaded semantic guidance for cross-view image translation,'' in \emph{Proceedings of the IEEE/CVF conference on computer vision and pattern recognition}, 2019, pp. 2412--2421.

\bibitem{Jiang2020TSITAS}
L.~Jiang, C.~Zhang, M.~Huang, C.~Liu, J.~Shi, and C.~C. Loy, ``Tsit: A simple and versatile framework for image-to-image translation,'' in \emph{Computer Vision--ECCV 2020: 16th European Conference, Glasgow, UK, August 23--28, 2020, Proceedings, Part III 16}.\hskip 1em plus 0.5em minus 0.4em\relax Springer, 2020, pp. 206--222.

\bibitem{Dundar2020PanopticBasedIS}
A.~Dundar, K.~Sapra, G.~Liu, A.~Tao, and B.~Catanzaro, ``Panoptic-based image synthesis,'' in \emph{Proceedings of the IEEE/CVF Conference on Computer Vision and Pattern Recognition}, 2020, pp. 8067--8076.

\bibitem{Tang2020DualAG}
H.~Tang, S.~Bai, and N.~Sebe, ``Dual attention gans for semantic image synthesis,'' in \emph{Proceedings of the 28th ACM International Conference on Multimedia}, 2020, pp. 1994--2002.

\bibitem{Zhu2020SemanticallyMI}
Z.~Zhu, Z.~Xu, A.~You, and X.~Bai, ``Semantically multi-modal image synthesis,'' in \emph{Proceedings of the IEEE/CVF conference on computer vision and pattern recognition}, 2020, pp. 5466--5475.

\bibitem{Lee2019MaskGANTD}
C.-H. Lee, Z.~Liu, L.~Wu, and P.~Luo, ``Maskgan: Towards diverse and interactive facial image manipulation,'' in \emph{Proceedings of the IEEE/CVF Conference on Computer Vision and Pattern Recognition}, 2019, pp. 5548--5557.

\bibitem{Zhu2019SEANIS}
P.~Zhu, R.~Abdal, Y.~Qin, and P.~Wonka, ``Sean: Image synthesis with semantic region-adaptive normalization,'' in \emph{Proceedings of the IEEE/CVF Conference on Computer Vision and Pattern Recognition}, 2019, pp. 5103--5112.

\bibitem{veeravasarapu2017adversarially}
V.~Veeravasarapu, C.~Rothkopf, and R.~Visvanathan, ``Adversarially tuned scene generation,'' in \emph{Proceedings of the IEEE Conference on Computer Vision and Pattern Recognition}, 2017, pp. 2587--2595.

\bibitem{li2023novel}
X.~Li, H.~Duan, B.~Liu, X.~Wang, and F.-Y. Wang, ``A novel framework to generate synthetic video for foreground detection in highway surveillance scenarios,'' \emph{IEEE Transactions on Intelligent Transportation Systems}, 2023.

\bibitem{Dosovitskiy2017CARLAAO}
A.~Dosovitskiy, G.~Ros, F.~Codevilla, A.~Lopez, and V.~Koltun, ``Carla: An open urban driving simulator,'' in \emph{Proceedings of the 1st Annual Conference on Robot Learning}.\hskip 1em plus 0.5em minus 0.4em\relax PMLR, 2017, pp. 1--16.

\bibitem{tian2023vistagpt}
Y.~Tian, X.~Li, H.~Zhang, C.~Zhao, B.~Li, X.~Wang, and F.-Y. Wang, ``Vistagpt: Generative parallel transformers for vehicles with intelligent systems for transport automation,'' \emph{IEEE Transactions on Intelligent Vehicles}, 2023.

\bibitem{kim2020reducing}
J.~Kim, J.~Ju, R.~Feldt, and S.~Yoo, ``Reducing dnn labelling cost using surprise adequacy: An industrial case study for autonomous driving,'' in \emph{Proceedings of the 28th ACM Joint Meeting on European Software Engineering Conference and Symposium on the Foundations of Software Engineering}, 2020, pp. 1466--1476.

\bibitem{song2023identifying}
R.~Song, X.~Li, X.~Zhao, M.~Liu, J.~Zhou, and F.-Y. Wang, ``Identifying critical test scenarios for lane keeping assistance system using analytic hierarchy process and hierarchical clustering,'' \emph{IEEE Transactions on Intelligent Vehicles}, 2023.

\bibitem{guo2023vectorized}
L.~Guo, C.~Shan, T.~Shi, X.~Li, and F.-Y. Wang, ``A vectorized representation model for trajectory prediction of intelligent vehicles in challenging scenarios,'' \emph{IEEE Transactions on Intelligent Vehicles}, 2023.

\bibitem{Ros2016TheSD}
G.~Ros, L.~Sellart, J.~Materzynska, D.~Vazquez, and A.~M. Lopez, ``The synthia dataset: A large collection of synthetic images for semantic segmentation of urban scenes,'' in \emph{Proceedings of the IEEE conference on computer vision and pattern recognition}, 2016, pp. 3234--3243.

\bibitem{saleh2018effective}
F.~S. Saleh, M.~S. Aliakbarian, M.~Salzmann, L.~Petersson, and J.~M. Alvarez, ``Effective use of synthetic data for urban scene semantic segmentation,'' in \emph{Proceedings of the European Conference on Computer Vision (ECCV)}, 2018, pp. 84--100.

\bibitem{Sakaridis2017SemanticFS}
C.~Sakaridis, D.~Dai, and L.~Van~Gool, ``Semantic foggy scene understanding with synthetic data,'' \emph{International Journal of Computer Vision}, vol. 126, pp. 973--992, 2018.

\bibitem{Alberti2020IDDAAL}
E.~Alberti, A.~Tavera, C.~Masone, and B.~Caputo, ``Idda: A large-scale multi-domain dataset for autonomous driving,'' \emph{IEEE Robotics and Automation Letters}, vol.~5, no.~4, pp. 5526--5533, 2020.

\bibitem{Kloukiniotis2022CarlaScenesAS}
A.~Kloukiniotis, A.~Papandreou, C.~Anagnostopoulos, A.~Lalos, P.~Kapsalas, D.-V. Nguyen, and K.~Moustakas, ``Carlascenes: A synthetic dataset for odometry in autonomous driving,'' in \emph{Proceedings of the IEEE/CVF Conference on Computer Vision and Pattern Recognition}, 2022, pp. 4519--4527.

\bibitem{agrawal2022comprehensive}
S.~C. Agrawal and A.~S. Jalal, ``A comprehensive review on analysis and implementation of recent image dehazing methods,'' \emph{Archives of Computational Methods in Engineering}, vol.~29, no.~7, pp. 4799--4850, 2022.

\bibitem{guo2023haze}
F.~Guo, J.~Yang, Z.~Liu, and J.~Tang, ``Haze removal for single image: A comprehensive review,'' \emph{Neurocomputing}, 2023.

\bibitem{shen2023optical}
S.~Shen, L.~Kerofsky, and S.~Yogamani, ``Optical flow for autonomous driving: Applications, challenges and improvements,'' \emph{arXiv preprint arXiv:2301.04422}, 2023.

\bibitem{kondermann2016hci}
D.~Kondermann, R.~Nair, K.~Honauer, K.~Krispin, J.~Andrulis, A.~Brock, B.~Gussefeld, M.~Rahimimoghaddam, S.~Hofmann, C.~Brenner \emph{et~al.}, ``The hci benchmark suite: Stereo and flow ground truth with uncertainties for urban autonomous driving,'' in \emph{Proceedings of the IEEE Conference on Computer Vision and Pattern Recognition Workshops}, 2016, pp. 19--28.

\bibitem{Wulff:ECCVws:2012}
J.~Wulff, D.~J. Butler, G.~B. Stanley, and M.~J. Black, ``Lessons and insights from creating a synthetic optical flow benchmark,'' in \emph{Computer Vision--ECCV 2012. Workshops and Demonstrations: Florence, Italy, October 7-13, 2012, Proceedings, Part II 12}.\hskip 1em plus 0.5em minus 0.4em\relax Springer, 2012, pp. 168--177.

\bibitem{sun2018pwc}
D.~Sun, X.~Yang, M.-Y. Liu, and J.~Kautz, ``Pwc-net: Cnns for optical flow using pyramid, warping, and cost volume,'' in \emph{Proceedings of the IEEE conference on computer vision and pattern recognition}, 2018, pp. 8934--8943.

\bibitem{liu2019selflow}
P.~Liu, M.~Lyu, I.~King, and J.~Xu, ``Selflow: Self-supervised learning of optical flow,'' in \emph{Proceedings of the IEEE/CVF conference on computer vision and pattern recognition}, 2019, pp. 4571--4580.

\bibitem{bar2020scopeflow}
A.~Bar-Haim and L.~Wolf, ``Scopeflow: Dynamic scene scoping for optical flow,'' in \emph{Proceedings of the IEEE/CVF Conference on Computer Vision and Pattern Recognition}, 2020, pp. 7998--8007.

\bibitem{hoyer2022mic}
L.~Hoyer, D.~Dai, H.~Wang, and L.~Van~Gool, ``Mic: Masked image consistency for context-enhanced domain adaptation,'' in \emph{Proceedings of the IEEE/CVF Conference on Computer Vision and Pattern Recognition}, 2023, pp. 11\,721--11\,732.

\bibitem{hoyer2022hrda}
L.~Hoyer, D.~Dai, and L.~Van~Gool, ``Hrda: Context-aware high-resolution domain-adaptive semantic segmentation,'' in \emph{European Conference on Computer Vision}.\hskip 1em plus 0.5em minus 0.4em\relax Springer, 2022, pp. 372--391.

\bibitem{xie2023sepico}
B.~Xie, S.~Li, M.~Li, C.~H. Liu, G.~Huang, and G.~Wang, ``Sepico: Semantic-guided pixel contrast for domain adaptive semantic segmentation,'' \emph{IEEE Transactions on Pattern Analysis and Machine Intelligence}, 2023.

\bibitem{Jurez2017SlantedSR}
D.~Hernandez-Juarez, L.~Schneider, A.~Espinosa, D.~V{\'a}zquez, A.~M. L{\'o}pez, U.~Franke, M.~Pollefeys, and J.~C.~M. Lopez, ``Slanted stixels: Representing san francisco’s steepest streets,'' in \emph{British Machine Vision Conference 2017, BMVC 2017}, vol.~87, 2017.

\bibitem{Bengar2019TemporalCF}
J.~Zolfaghari~Bengar, A.~Gonzalez-Garcia, G.~Villalonga, B.~Raducanu, H.~Habibi~Aghdam, M.~Mozerov, A.~M. Lopez, and J.~Van~de Weijer, ``Temporal coherence for active learning in videos,'' in \emph{Proceedings of the IEEE/CVF International Conference on Computer Vision Workshops}, 2019, pp. 0--0.

\bibitem{Cordts2016Cityscapes}
M.~Cordts, M.~Omran, S.~Ramos, T.~Rehfeld, M.~Enzweiler, R.~Benenson, U.~Franke, S.~Roth, and B.~Schiele, ``The cityscapes dataset for semantic urban scene understanding,'' in \emph{Proceedings of the IEEE conference on computer vision and pattern recognition}, 2016, pp. 3213--3223.

\bibitem{mandal2020real}
G.~Mandal, P.~De, and D.~Bhattacharya, ``A real-time fast defogging system to clear the vision of driver in foggy highway using minimum filter and gamma correction,'' \emph{S{\=a}dhan{\=a}}, vol.~45, pp. 1--11, 2020.

\bibitem{gadipudi2022synthetic}
N.~Gadipudi, I.~Elamvazuthi, M.~Sanmugam, L.~I. Izhar, T.~Prasetyo, R.~Jegadeeshwaran, and S.~S.~A. Ali, ``Synthetic to real gap estimation of autonomous driving datasets using feature embedding,'' in \emph{2022 IEEE 5th International Symposium in Robotics and Manufacturing Automation (ROMA)}.\hskip 1em plus 0.5em minus 0.4em\relax IEEE, 2022, pp. 1--5.

\bibitem{engel2017direct}
J.~Engel, V.~Koltun, and D.~Cremers, ``Direct sparse odometry,'' \emph{IEEE transactions on pattern analysis and machine intelligence}, vol.~40, no.~3, pp. 611--625, 2017.

\bibitem{shan2018lego}
T.~Shan and B.~Englot, ``Lego-loam: Lightweight and ground-optimized lidar odometry and mapping on variable terrain,'' in \emph{2018 IEEE/RSJ International Conference on Intelligent Robots and Systems (IROS)}.\hskip 1em plus 0.5em minus 0.4em\relax IEEE, 2018, pp. 4758--4765.

\bibitem{yang2018deep}
N.~Yang, R.~Wang, J.~Stuckler, and D.~Cremers, ``Deep virtual stereo odometry: Leveraging deep depth prediction for monocular direct sparse odometry,'' in \emph{Proceedings of the European conference on computer vision (ECCV)}, 2018, pp. 817--833.

\bibitem{Richter2017PlayingFB}
S.~R. Richter, Z.~Hayder, and V.~Koltun, ``Playing for benchmarks,'' in \emph{Proceedings of the IEEE International Conference on Computer Vision}, 2017, pp. 2232--2241.

\bibitem{Hurl2019PreciseSI}
B.~Hurl, K.~Czarnecki, and S.~Waslander, ``Precise synthetic image and lidar (presil) dataset for autonomous vehicle perception,'' in \emph{2019 IEEE Intelligent Vehicles Symposium (IV)}.\hskip 1em plus 0.5em minus 0.4em\relax IEEE, 2019, pp. 2522--2529.

\bibitem{li2018paralleleye}
X.~Li, K.~Wang, Y.~Tian, L.~Yan, F.~Deng, and F.-Y. Wang, ``The paralleleye dataset: A large collection of virtual images for traffic vision research,'' \emph{IEEE Transactions on Intelligent Transportation Systems}, vol.~20, no.~6, pp. 2072--2084, 2018.

\bibitem{Cabon2020VirtualK2}
Y.~Cabon, N.~Murray, and M.~Humenberger, ``Virtual kitti 2,'' \emph{arXiv preprint arXiv:2001.10773}, 2020.

\bibitem{Sun2022SHIFTAS}
T.~Sun, M.~Segu, J.~Postels, Y.~Wang, L.~Van~Gool, B.~Schiele, F.~Tombari, and F.~Yu, ``Shift: a synthetic driving dataset for continuous multi-task domain adaptation,'' in \emph{Proceedings of the IEEE/CVF Conference on Computer Vision and Pattern Recognition}, 2022, pp. 21\,339--21\,350.

\bibitem{Li2022V2XSimMC}
Y.~Li, D.~Ma, Z.~An, Z.~Wang, Y.~Zhong, S.~Chen, and C.~Feng, ``V2x-sim: Multi-agent collaborative perception dataset and benchmark for autonomous driving,'' \emph{IEEE Robotics and Automation Letters}, vol.~7, no.~4, pp. 10\,914--10\,921, 2022.

\bibitem{AIODrive}
X.~Weng, Y.~Man, D.~Cheng, J.~Park, M.~O’Toole, K.~Kitani, J.~Wang, and D.~Held, ``All-in-one drive: A large-scale comprehensive perception dataset with high-density long-range point clouds,'' \emph{arXiv}, 2020.

\bibitem{Xu2021OPV2VAO}
R.~Xu, H.~Xiang, X.~Xia, X.~Han, J.~Li, and J.~Ma, ``Opv2v: An open benchmark dataset and fusion pipeline for perception with vehicle-to-vehicle communication,'' in \emph{2022 International Conference on Robotics and Automation (ICRA)}.\hskip 1em plus 0.5em minus 0.4em\relax IEEE, 2022, pp. 2583--2589.

\bibitem{zhang2020virtual}
H.~Zhang, G.~Luo, Y.~Tian, K.~Wang, H.~He, and F.-Y. Wang, ``A virtual-real interaction approach to object instance segmentation in traffic scenes,'' \emph{IEEE Transactions on Intelligent Transportation Systems}, vol.~22, no.~2, pp. 863--875, 2020.

\bibitem{chen2019learning}
Y.~Chen, W.~Li, X.~Chen, and L.~V. Gool, ``Learning semantic segmentation from synthetic data: A geometrically guided input-output adaptation approach,'' in \emph{Proceedings of the IEEE/CVF conference on computer vision and pattern recognition}, 2019, pp. 1841--1850.

\bibitem{su2020adapting}
P.~Su, K.~Wang, X.~Zeng, S.~Tang, D.~Chen, D.~Qiu, and X.~Wang, ``Adapting object detectors with conditional domain normalization,'' in \emph{Computer Vision--ECCV 2020: 16th European Conference, Glasgow, UK, August 23--28, 2020, Proceedings, Part XI 16}.\hskip 1em plus 0.5em minus 0.4em\relax Springer, 2020, pp. 403--419.

\bibitem{wrenninge2018synscapes}
M.~Wrenninge and J.~Unger, ``Synscapes: A photorealistic synthetic dataset for street scene parsing,'' \emph{arXiv preprint arXiv:1810.08705}, 2018.

\bibitem{johnson2016driving}
M.~Johnson-Roberson, C.~Barto, R.~Mehta, S.~N. Sridhar, K.~Rosaen, and R.~Vasudevan, ``Driving in the matrix: Can virtual worlds replace human-generated annotations for real world tasks?'' \emph{arXiv preprint arXiv:1610.01983}, 2016.

\bibitem{li2023paralleleye}
X.~Li, K.~Wang, X.~Gu, F.~Deng, and F.-Y. Wang, ``Paralleleye pipeline: An effective method to synthesize images for improving the visual intelligence of intelligent vehicles,'' \emph{IEEE Transactions on Systems, Man, and Cybernetics: Systems}, 2023.

\bibitem{li2019paralleleye}
X.~Li, Y.~Wang, L.~Yan, K.~Wang, F.~Deng, and F.-Y. Wang, ``Paralleleye-cs: A new dataset of synthetic images for testing the visual intelligence of intelligent vehicles,'' \emph{IEEE Transactions on Vehicular Technology}, vol.~68, no.~10, pp. 9619--9631, 2019.

\bibitem{ancha2020active}
S.~Ancha, Y.~Raaj, P.~Hu, S.~G. Narasimhan, and D.~Held, ``Active perception using light curtains for autonomous driving,'' in \emph{Computer Vision--ECCV 2020: 16th European Conference, Glasgow, UK, August 23--28, 2020, Proceedings, Part V 16}.\hskip 1em plus 0.5em minus 0.4em\relax Springer, 2020, pp. 751--766.

\bibitem{qu2020depth}
C.~Qu, T.~Nguyen, and C.~Taylor, ``Depth completion via deep basis fitting,'' in \emph{Proceedings of the IEEE/CVF Winter Conference on Applications of Computer Vision}, 2020, pp. 71--80.

\bibitem{lagos2022pandepth}
J.~Lagos and E.~Rahtu, ``Pandepth: Joint panoptic segmentation and depth completion,'' \emph{arXiv preprint arXiv:2212.14180}, 2022.

\bibitem{jaritz2022cross}
M.~Jaritz, T.-H. Vu, R.~De~Charette, {\'E}.~Wirbel, and P.~P{\'e}rez, ``Cross-modal learning for domain adaptation in 3d semantic segmentation,'' \emph{IEEE Transactions on Pattern Analysis and Machine Intelligence}, vol.~45, no.~2, pp. 1533--1544, 2022.

\bibitem{lei2022latency}
Z.~Lei, S.~Ren, Y.~Hu, W.~Zhang, and S.~Chen, ``Latency-aware collaborative perception,'' in \emph{Computer Vision--ECCV 2022: 17th European Conference, Tel Aviv, Israel, October 23--27, 2022, Proceedings, Part XXXII}.\hskip 1em plus 0.5em minus 0.4em\relax Springer, 2022, pp. 316--332.

\bibitem{lu2022robust}
Y.~Lu, Q.~Li, B.~Liu, M.~Dianati, C.~Feng, S.~Chen, and Y.~Wang, ``Robust collaborative 3d object detection in presence of pose errors,'' in \emph{2023 IEEE International Conference on Robotics and Automation (ICRA)}.\hskip 1em plus 0.5em minus 0.4em\relax IEEE, 2023, pp. 4812--4818.

\bibitem{hu2022where2comm}
Y.~Hu, S.~Fang, Z.~Lei, Y.~Zhong, and S.~Chen, ``Where2comm: Communication-efficient collaborative perception via spatial confidence maps,'' \emph{Advances in neural information processing systems}, vol.~35, pp. 4874--4886, 2022.

\bibitem{yu2022dair}
H.~Yu, Y.~Luo, M.~Shu, Y.~Huo, Z.~Yang, Y.~Shi, Z.~Guo, H.~Li, X.~Hu, J.~Yuan \emph{et~al.}, ``Dair-v2x: A large-scale dataset for vehicle-infrastructure cooperative 3d object detection,'' in \emph{Proceedings of the IEEE/CVF Conference on Computer Vision and Pattern Recognition}, 2022, pp. 21\,361--21\,370.

\bibitem{geirhos2020shortcut}
R.~Geirhos, J.-H. Jacobsen, C.~Michaelis, R.~Zemel, W.~Brendel, M.~Bethge, and F.~A. Wichmann, ``Shortcut learning in deep neural networks,'' \emph{Nature Machine Intelligence}, vol.~2, no.~11, pp. 665--673, 2020.

\bibitem{jain2021autonomy}
A.~Jain, L.~Del~Pero, H.~Grimmett, and P.~Ondruska, ``Autonomy 2.0: Why is self-driving always 5 years away?'' \emph{arXiv preprint arXiv:2107.08142}, 2021.

\bibitem{makansi2021exposing}
O.~Makansi, {\"O}.~Cicek, Y.~Marrakchi, and T.~Brox, ``On exposing the challenging long tail in future prediction of traffic actors,'' in \emph{Proceedings of the IEEE/CVF International Conference on Computer Vision}, 2021, pp. 13\,147--13\,157.

\bibitem{liang2022advances}
W.~Liang, G.~A. Tadesse, D.~Ho, L.~Fei-Fei, M.~Zaharia, C.~Zhang, and J.~Zou, ``Advances, challenges and opportunities in creating data for trustworthy ai,'' \emph{Nature Machine Intelligence}, vol.~4, no.~8, pp. 669--677, 2022.

\bibitem{northcutt2021confident}
C.~Northcutt, L.~Jiang, and I.~Chuang, ``Confident learning: Estimating uncertainty in dataset labels,'' \emph{Journal of Artificial Intelligence Research}, vol.~70, pp. 1373--1411, 2021.

\bibitem{northcutt2021pervasive}
C.~G. Northcutt, A.~Athalye, and J.~Mueller, ``Pervasive label errors in test sets destabilize machine learning benchmarks,'' \emph{arXiv preprint arXiv:2103.14749}, 2021.

\bibitem{van2001art}
D.~A. Van~Dyk and X.-L. Meng, ``The art of data augmentation,'' \emph{Journal of Computational and Graphical Statistics}, vol.~10, no.~1, pp. 1--50, 2001.

\bibitem{antoniou2017data}
A.~Antoniou, A.~Storkey, and H.~Edwards, ``Data augmentation generative adversarial networks,'' \emph{arXiv preprint arXiv:1711.04340}, 2017.

\bibitem{liang2021neural}
W.~Liang and J.~Zou, ``Neural group testing to accelerate deep learning,'' in \emph{2021 IEEE International Symposium on Information Theory (ISIT)}.\hskip 1em plus 0.5em minus 0.4em\relax IEEE, 2021, pp. 958--963.

\bibitem{ros2016training}
G.~Ros, S.~Stent, P.~F. Alcantarilla, and T.~Watanabe, ``Training constrained deconvolutional networks for road scene semantic segmentation,'' \emph{arXiv preprint arXiv:1604.01545}, 2016.

\bibitem{Wang2018DeepVD}
M.~Wang and W.~Deng, ``Deep visual domain adaptation: A survey,'' \emph{Neurocomputing}, vol. 312, pp. 135--153, 2018.

\bibitem{Li2017DeeperBA}
D.~Li, Y.~Yang, Y.-Z. Song, and T.~M. Hospedales, ``Deeper, broader and artier domain generalization,'' in \emph{Proceedings of the IEEE international conference on computer vision}, 2017, pp. 5543--5551.

\bibitem{Segu2020BatchNE}
M.~Segu, A.~Tonioni, and F.~Tombari, ``Batch normalization embeddings for deep domain generalization,'' \emph{Pattern Recognit.}, vol. 135, p. 109115, 2020.

\bibitem{Volpi2020ContinualAO}
R.~Volpi, D.~Larlus, and G.~Rogez, ``Continual adaptation of visual representations via domain randomization and meta-learning,'' in \emph{Proceedings of the IEEE/CVF Conference on Computer Vision and Pattern Recognition}, 2021, pp. 4441--4451.

\bibitem{Wang2021TentFT}
D.~Wang, E.~Shelhamer, S.~Liu, B.~A. Olshausen, and T.~Darrell, ``Tent: Fully test-time adaptation by entropy minimization,'' in \emph{International Conference on Learning Representations}, 2021.

\bibitem{guo2017calibration}
C.~Guo, G.~Pleiss, Y.~Sun, and K.~Q. Weinberger, ``On calibration of modern neural networks,'' in \emph{International conference on machine learning}.\hskip 1em plus 0.5em minus 0.4em\relax PMLR, 2017, pp. 1321--1330.

\bibitem{gal2016dropout}
Y.~Gal and Z.~Ghahramani, ``Dropout as a bayesian approximation: Representing model uncertainty in deep learning,'' in \emph{international conference on machine learning}.\hskip 1em plus 0.5em minus 0.4em\relax PMLR, 2016, pp. 1050--1059.

\bibitem{Lakshminarayanan2016SimpleAS}
B.~Lakshminarayanan, A.~Pritzel, and C.~Blundell, ``Simple and scalable predictive uncertainty estimation using deep ensembles,'' \emph{Advances in neural information processing systems}, vol.~30, 2017.

\bibitem{prakash2021self}
A.~Prakash, S.~Debnath, J.-F. Lafleche, E.~Cameracci, S.~Birchfield, M.~T. Law \emph{et~al.}, ``Self-supervised real-to-sim scene generation,'' in \emph{Proceedings of the IEEE/CVF International Conference on Computer Vision}, 2021, pp. 16\,044--16\,054.

\bibitem{prabhu2023bridging}
V.~Prabhu, D.~Acuna, A.~Liao, R.~Mahmood, M.~T. Law, J.~Hoffman, S.~Fidler, and J.~Lucas, ``Bridging the sim2real gap with care: Supervised detection adaptation with conditional alignment and reweighting,'' \emph{arXiv preprint arXiv:2302.04832}, 2023.

\bibitem{tobin2017domain}
J.~Tobin, R.~Fong, A.~Ray, J.~Schneider, W.~Zaremba, and P.~Abbeel, ``Domain randomization for transferring deep neural networks from simulation to the real world,'' in \emph{2017 IEEE/RSJ international conference on intelligent robots and systems (IROS)}.\hskip 1em plus 0.5em minus 0.4em\relax IEEE, 2017, pp. 23--30.

\bibitem{tremblay2018training}
J.~Tremblay, A.~Prakash, D.~Acuna, M.~Brophy, V.~Jampani, C.~Anil, T.~To, E.~Cameracci, S.~Boochoon, and S.~Birchfield, ``Training deep networks with synthetic data: Bridging the reality gap by domain randomization,'' in \emph{Proceedings of the IEEE conference on computer vision and pattern recognition workshops}, 2018, pp. 969--977.

\bibitem{prakash2019structured}
A.~Prakash, S.~Boochoon, M.~Brophy, D.~Acuna, E.~Cameracci, G.~State, O.~Shapira, and S.~Birchfield, ``Structured domain randomization: Bridging the reality gap by context-aware synthetic data,'' in \emph{2019 International Conference on Robotics and Automation (ICRA)}.\hskip 1em plus 0.5em minus 0.4em\relax IEEE, 2019, pp. 7249--7255.

\bibitem{huang2018multimodal}
X.~Huang, M.-Y. Liu, S.~Belongie, and J.~Kautz, ``Multimodal unsupervised image-to-image translation,'' in \emph{Proceedings of the European conference on computer vision (ECCV)}, 2018, pp. 172--189.

\bibitem{hoffman2018cycada}
J.~Hoffman, E.~Tzeng, T.~Park, J.-Y. Zhu, P.~Isola, K.~Saenko, A.~Efros, and T.~Darrell, ``Cycada: Cycle-consistent adversarial domain adaptation,'' in \emph{International conference on machine learning}.\hskip 1em plus 0.5em minus 0.4em\relax Pmlr, 2018, pp. 1989--1998.

\bibitem{chen2019crdoco}
Y.-C. Chen, Y.-Y. Lin, M.-H. Yang, and J.-B. Huang, ``Crdoco: Pixel-level domain transfer with cross-domain consistency,'' in \emph{Proceedings of the IEEE/CVF conference on computer vision and pattern recognition}, 2019, pp. 1791--1800.

\bibitem{Tsai2018LearningTA}
Y.-H. Tsai, W.-C. Hung, S.~Schulter, K.~Sohn, M.-H. Yang, and M.~Chandraker, ``Learning to adapt structured output space for semantic segmentation,'' in \emph{Proceedings of the IEEE conference on computer vision and pattern recognition}, 2018, pp. 7472--7481.

\bibitem{saito2019strong}
K.~Saito, Y.~Ushiku, T.~Harada, and K.~Saenko, ``Strong-weak distribution alignment for adaptive object detection,'' in \emph{Proceedings of the IEEE/CVF Conference on Computer Vision and Pattern Recognition}, 2019, pp. 6956--6965.

\bibitem{zhu2019adapting}
X.~Zhu, J.~Pang, C.~Yang, J.~Shi, and D.~Lin, ``Adapting object detectors via selective cross-domain alignment,'' in \emph{Proceedings of the IEEE/CVF Conference on Computer Vision and Pattern Recognition}, 2019, pp. 687--696.

\bibitem{hsu2020progressive}
H.-K. Hsu, C.-H. Yao, Y.-H. Tsai, W.-C. Hung, H.-Y. Tseng, M.~Singh, and M.-H. Yang, ``Progressive domain adaptation for object detection,'' in \emph{Proceedings of the IEEE/CVF winter conference on applications of computer vision}, 2020, pp. 749--757.

\bibitem{yu2022sc}
F.~Yu, D.~Wang, Y.~Chen, N.~Karianakis, T.~Shen, P.~Yu, D.~Lymberopoulos, S.~Lu, W.~Shi, and X.~Chen, ``Sc-uda: Style and content gaps aware unsupervised domain adaptation for object detection,'' in \emph{Proceedings of the IEEE/CVF Winter Conference on Applications of Computer Vision}, 2022, pp. 382--391.

\bibitem{tan2020class}
S.~Tan, X.~Peng, and K.~Saenko, ``Class-imbalanced domain adaptation: an empirical odyssey,'' in \emph{Computer Vision--ECCV 2020 Workshops: Glasgow, UK, August 23--28, 2020, Proceedings, Part I 16}.\hskip 1em plus 0.5em minus 0.4em\relax Springer, 2020, pp. 585--602.

\bibitem{li2019bidirectional}
Y.~Li, L.~Yuan, and N.~Vasconcelos, ``Bidirectional learning for domain adaptation of semantic segmentation,'' in \emph{Proceedings of the IEEE/CVF Conference on Computer Vision and Pattern Recognition}, 2019, pp. 6936--6945.

\bibitem{Kar2019MetaSimLT}
A.~Kar, A.~Prakash, M.-Y. Liu, E.~Cameracci, J.~Yuan, M.~Rusiniak, D.~Acuna, A.~Torralba, and S.~Fidler, ``Meta-sim: Learning to generate synthetic datasets,'' in \emph{Proceedings of the IEEE/CVF International Conference on Computer Vision}, 2019, pp. 4551--4560.

\bibitem{devaranjan2020meta}
J.~Devaranjan, A.~Kar, and S.~Fidler, ``Meta-sim2: Unsupervised learning of scene structure for synthetic data generation,'' in \emph{Computer Vision--ECCV 2020: 16th European Conference, Glasgow, UK, August 23--28, 2020, Proceedings, Part XVII 16}.\hskip 1em plus 0.5em minus 0.4em\relax Springer, 2020, pp. 715--733.

\bibitem{tan2021scenegen}
S.~Tan, K.~Wong, S.~Wang, S.~Manivasagam, M.~Ren, and R.~Urtasun, ``Scenegen: Learning to generate realistic traffic scenes,'' in \emph{Proceedings of the IEEE/CVF Conference on Computer Vision and Pattern Recognition}, 2021, pp. 892--901.

\bibitem{mildenhall2021nerf}
B.~Mildenhall, P.~P. Srinivasan, M.~Tancik, J.~T. Barron, R.~Ramamoorthi, and R.~Ng, ``Nerf: Representing scenes as neural radiance fields for view synthesis,'' \emph{Communications of the ACM}, vol.~65, no.~1, pp. 99--106, 2021.

\bibitem{zhang2020nerf++}
K.~Zhang, G.~Riegler, N.~Snavely, and V.~Koltun, ``Nerf++: Analyzing and improving neural radiance fields,'' \emph{arXiv preprint arXiv:2010.07492}, 2020.

\bibitem{muller2022instant}
T.~M{\"u}ller, A.~Evans, C.~Schied, and A.~Keller, ``Instant neural graphics primitives with a multiresolution hash encoding,'' \emph{ACM Transactions on Graphics (ToG)}, vol.~41, no.~4, pp. 1--15, 2022.

\bibitem{tancik2022block}
M.~Tancik, V.~Casser, X.~Yan, S.~Pradhan, B.~Mildenhall, P.~P. Srinivasan, J.~T. Barron, and H.~Kretzschmar, ``Block-nerf: Scalable large scene neural view synthesis,'' in \emph{Proceedings of the IEEE/CVF Conference on Computer Vision and Pattern Recognition}, 2022, pp. 8248--8258.

\bibitem{chen2021mvsnerf}
A.~Chen, Z.~Xu, F.~Zhao, X.~Zhang, F.~Xiang, J.~Yu, and H.~Su, ``Mvsnerf: Fast generalizable radiance field reconstruction from multi-view stereo,'' in \emph{Proceedings of the IEEE/CVF International Conference on Computer Vision}, 2021, pp. 14\,124--14\,133.

\bibitem{song2017semantic}
S.~Song, F.~Yu, A.~Zeng, A.~X. Chang, M.~Savva, and T.~Funkhouser, ``Semantic scene completion from a single depth image,'' in \emph{Proceedings of the IEEE conference on computer vision and pattern recognition}, 2017, pp. 1746--1754.

\bibitem{li2019rgbd}
J.~Li, Y.~Liu, D.~Gong, Q.~Shi, X.~Yuan, C.~Zhao, and I.~Reid, ``Rgbd based dimensional decomposition residual network for 3d semantic scene completion,'' in \emph{Proceedings of the IEEE/CVF Conference on Computer Vision and Pattern Recognition}, 2019, pp. 7693--7702.

\bibitem{li2019depth}
J.~Li, Y.~Liu, X.~Yuan, C.~Zhao, R.~Siegwart, I.~Reid, and C.~Cadena, ``Depth based semantic scene completion with position importance aware loss,'' \emph{IEEE Robotics and Automation Letters}, vol.~5, no.~1, pp. 219--226, 2019.

\bibitem{li2020anisotropic}
J.~Li, K.~Han, P.~Wang, Y.~Liu, and X.~Yuan, ``Anisotropic convolutional networks for 3d semantic scene completion,'' in \emph{Proceedings of the IEEE/CVF Conference on Computer Vision and Pattern Recognition}, 2020, pp. 3351--3359.

\bibitem{cheng2021s3cnet}
R.~Cheng, C.~Agia, Y.~Ren, X.~Li, and L.~Bingbing, ``S3cnet: A sparse semantic scene completion network for lidar point clouds,'' in \emph{Conference on Robot Learning}.\hskip 1em plus 0.5em minus 0.4em\relax PMLR, 2021, pp. 2148--2161.

\bibitem{firman2016structured}
M.~Firman, O.~Mac~Aodha, S.~Julier, and G.~J. Brostow, ``Structured prediction of unobserved voxels from a single depth image,'' in \emph{Proceedings of the IEEE Conference on Computer Vision and Pattern Recognition}, 2016, pp. 5431--5440.

\bibitem{xu2022scene}
M.~Xu, P.~Chen, H.~Liu, and X.~Han, ``To-scene: A large-scale dataset for understanding 3d tabletop scenes,'' in \emph{Computer Vision--ECCV 2022: 17th European Conference, Tel Aviv, Israel, October 23--27, 2022, Proceedings, Part XXVII}.\hskip 1em plus 0.5em minus 0.4em\relax Springer, 2022, pp. 340--356.

\bibitem{chang2015shapenet}
A.~X. Chang, T.~Funkhouser, L.~Guibas, P.~Hanrahan, Q.~Huang, Z.~Li, S.~Savarese, M.~Savva, S.~Song, H.~Su \emph{et~al.}, ``Shapenet: An information-rich 3d model repository,'' \emph{arXiv preprint arXiv:1512.03012}, 2015.

\bibitem{curless1996volumetric}
B.~Curless and M.~Levoy, ``A volumetric method for building complex models from range images,'' in \emph{Proceedings of the 23rd annual conference on Computer graphics and interactive techniques}, 1996, pp. 303--312.

\bibitem{loper2015smpl}
M.~Loper, N.~Mahmood, J.~Romero, G.~Pons-Moll, and M.~J. Black, ``Smpl: A skinned multi-person linear model,'' \emph{ACM transactions on graphics (TOG)}, vol.~34, no.~6, pp. 1--16, 2015.

\bibitem{saito2019pifu}
S.~Saito, Z.~Huang, R.~Natsume, S.~Morishima, A.~Kanazawa, and H.~Li, ``Pifu: Pixel-aligned implicit function for high-resolution clothed human digitization,'' in \emph{Proceedings of the IEEE/CVF international conference on computer vision}, 2019, pp. 2304--2314.

\bibitem{national2019national}
N.~Science and T.~C. U. S.~C. on~Artificial~Intelligence, \emph{The national artificial intelligence research and development strategic plan: 2019 update}.\hskip 1em plus 0.5em minus 0.4em\relax National Science and Technology Council (US), Select Committee on Artificial~…, 2019.

\bibitem{xu2022safebench}
C.~Xu, W.~Ding, W.~Lyu, Z.~Liu, S.~Wang, Y.~He, H.~Hu, D.~Zhao, and B.~Li, ``Safebench: A benchmarking platform for safety evaluation of autonomous vehicles,'' \emph{Advances in Neural Information Processing Systems}, vol.~35, pp. 25\,667--25\,682, 2022.

\bibitem{ghorbani2019data}
A.~Ghorbani and J.~Zou, ``Data shapley: Equitable valuation of data for machine learning,'' in \emph{International Conference on Machine Learning}.\hskip 1em plus 0.5em minus 0.4em\relax PMLR, 2019, pp. 2242--2251.

\bibitem{kwon2021efficient}
Y.~Kwon, M.~A. Rivas, and J.~Zou, ``Efficient computation and analysis of distributional shapley values,'' in \emph{International Conference on Artificial Intelligence and Statistics}.\hskip 1em plus 0.5em minus 0.4em\relax PMLR, 2021, pp. 793--801.

\bibitem{jia2019towards}
R.~Jia, D.~Dao, B.~Wang, F.~A. Hubis, N.~Hynes, N.~M. G{\"u}rel, B.~Li, C.~Zhang, D.~Song, and C.~J. Spanos, ``Towards efficient data valuation based on the shapley value,'' in \emph{The 22nd International Conference on Artificial Intelligence and Statistics}.\hskip 1em plus 0.5em minus 0.4em\relax PMLR, 2019, pp. 1167--1176.

\bibitem{koh2017understanding}
P.~W. Koh and P.~Liang, ``Understanding black-box predictions via influence functions,'' in \emph{International conference on machine learning}.\hskip 1em plus 0.5em minus 0.4em\relax PMLR, 2017, pp. 1885--1894.

\bibitem{li2022features}
X.~Li, P.~Ye, J.~Li, Z.~Liu, L.~Cao, and F.-Y. Wang, ``From features engineering to scenarios engineering for trustworthy ai: I\&i, c\&c, and v\&v,'' \emph{IEEE Intelligent Systems}, vol.~37, no.~4, pp. 18--26, 2022.

\bibitem{li2022novel}
X.~Li, Y.~Tian, P.~Ye, H.~Duan, and F.-Y. Wang, ``A novel scenarios engineering methodology for foundation models in metaverse,'' \emph{IEEE Transactions on Systems, Man, and Cybernetics: Systems}, vol.~53, no.~4, pp. 2148--2159, 2022.

\bibitem{california}
\BIBentryALTinterwordspacing
California department of motor vehicle disengagement report. (2021). [Online]. Available: \url{https://www.dmv.ca.gov/portal/vehicle-industry-services/autonomous-vehicles/disengagement-reports/}
\BIBentrySTDinterwordspacing

\bibitem{zhao2016accelerated}
D.~Zhao, ``Accelerated evaluation of automated vehicles.'' Ph.D. dissertation, 2016.

\bibitem{o2018scalable}
M.~O'Kelly, A.~Sinha, H.~Namkoong, R.~Tedrake, and J.~C. Duchi, ``Scalable end-to-end autonomous vehicle testing via rare-event simulation,'' \emph{Advances in neural information processing systems}, vol.~31, 2018.

\bibitem{wallace2011class}
B.~C. Wallace, K.~Small, C.~E. Brodley, and T.~A. Trikalinos, ``Class imbalance, redux,'' in \emph{2011 IEEE 11th international conference on data mining}.\hskip 1em plus 0.5em minus 0.4em\relax Ieee, 2011, pp. 754--763.

\bibitem{arief2021deep}
M.~Arief, Z.~Huang, G.~K.~S. Kumar, Y.~Bai, S.~He, W.~Ding, H.~Lam, and D.~Zhao, ``Deep probabilistic accelerated evaluation: A robust certifiable rare-event simulation methodology for black-box safety-critical systems,'' in \emph{International Conference on Artificial Intelligence and Statistics}.\hskip 1em plus 0.5em minus 0.4em\relax PMLR, 2021, pp. 595--603.

\bibitem{menon2020long}
A.~K. Menon, S.~Jayasumana, A.~S. Rawat, H.~Jain, A.~Veit, and S.~Kumar, ``Long-tail learning via logit adjustment,'' \emph{arXiv preprint arXiv:2007.07314}, 2020.

\bibitem{kim2020adjusting}
B.~Kim and J.~Kim, ``Adjusting decision boundary for class imbalanced learning,'' \emph{IEEE Access}, vol.~8, pp. 81\,674--81\,685, 2020.

\bibitem{mao2021one}
J.~Mao, M.~Niu, C.~Jiang, H.~Liang, J.~Chen, X.~Liang, Y.~Li, C.~Ye, W.~Zhang, Z.~Li \emph{et~al.}, ``One million scenes for autonomous driving: Once dataset,'' \emph{arXiv preprint arXiv:2106.11037}, 2021.

\bibitem{wilson2023argoverse}
B.~Wilson, W.~Qi, T.~Agarwal, J.~Lambert, J.~Singh, S.~Khandelwal, B.~Pan, R.~Kumar, A.~Hartnett, J.~K. Pontes \emph{et~al.}, ``Argoverse 2: Next generation datasets for self-driving perception and forecasting,'' \emph{arXiv preprint arXiv:2301.00493}, 2023.

\bibitem{ding2021multimodal}
W.~Ding, B.~Chen, B.~Li, K.~J. Eun, and D.~Zhao, ``Multimodal safety-critical scenarios generation for decision-making algorithms evaluation,'' \emph{IEEE Robotics and Automation Letters}, vol.~6, no.~2, pp. 1551--1558, 2021.

\bibitem{zhang2022adversarial}
Q.~Zhang, S.~Hu, J.~Sun, Q.~A. Chen, and Z.~M. Mao, ``On adversarial robustness of trajectory prediction for autonomous vehicles,'' in \emph{Proceedings of the IEEE/CVF Conference on Computer Vision and Pattern Recognition}, 2022, pp. 15\,159--15\,168.

\bibitem{feng2021intelligent}
S.~Feng, X.~Yan, H.~Sun, Y.~Feng, and H.~X. Liu, ``Intelligent driving intelligence test for autonomous vehicles with naturalistic and adversarial environment,'' \emph{Nature communications}, vol.~12, no.~1, p. 748, 2021.

\bibitem{ding2021semantically}
W.~Ding, H.~Lin, B.~Li, K.~J. Eun, and D.~Zhao, ``Semantically adversarial driving scenario generation with explicit knowledge integration,'' \emph{arXiv preprint arXiv:2106.04066}, 2021.

\bibitem{bagschik2018ontology}
G.~Bagschik, T.~Menzel, and M.~Maurer, ``Ontology based scene creation for the development of automated vehicles,'' in \emph{2018 IEEE Intelligent Vehicles Symposium (IV)}.\hskip 1em plus 0.5em minus 0.4em\relax IEEE, 2018, pp. 1813--1820.

\bibitem{saaty2006decision}
T.~L. Saaty, L.~G. Vargas \emph{et~al.}, \emph{Decision making with the analytic network process}.\hskip 1em plus 0.5em minus 0.4em\relax Springer, 2006, vol. 282.

\bibitem{saaty2013analytic}
T.~L. Saaty, L.~G. Vargas, T.~L. Saaty, and L.~G. Vargas, \emph{The analytic network process}.\hskip 1em plus 0.5em minus 0.4em\relax Springer, 2013.

\bibitem{hore2010image}
A.~Hore and D.~Ziou, ``Image quality metrics: Psnr vs. ssim,'' in \emph{2010 20th international conference on pattern recognition}.\hskip 1em plus 0.5em minus 0.4em\relax IEEE, 2010, pp. 2366--2369.

\bibitem{saaty1988analytic}
T.~L. Saaty, \emph{What is the analytic hierarchy process?}\hskip 1em plus 0.5em minus 0.4em\relax Springer, 1988.

\bibitem{saaty2008decision}
------, ``Decision making with the analytic hierarchy process,'' \emph{International journal of services sciences}, vol.~1, no.~1, pp. 83--98, 2008.

\bibitem{adams2003super}
W.~Adams and R.~Saaty, ``Super decisions software guide,'' \emph{Super Decisions}, vol.~9, p.~43, 2003.

\bibitem{yu2023benchmarking}
K.~Yu, T.~Tao, H.~Xie, Z.~Lin, T.~Liang, B.~Wang, P.~Chen, D.~Hao, Y.~Wang, and X.~Liang, ``Benchmarking the robustness of lidar-camera fusion for 3d object detection,'' in \emph{Proceedings of the IEEE/CVF Conference on Computer Vision and Pattern Recognition}, 2023, pp. 3187--3197.

\end{thebibliography}

% can use a bibliography generated by BibTeX as a .bbl file
% BibTeX documentation can be easily obtained at:
% http://mirror.ctan.org/biblio/bibtex/contrib/doc/
% The IEEEtran BibTeX style support page is at:
% http://www.michaelshell.org/tex/ieeetran/bibtex/
%\bibliographystyle{IEEEtran}
% argument is your BibTeX string definitions and bibliography database(s)
%\bibliography{IEEEabrv,../bib/paper}
%
% <OR> manually copy in the resultant .bbl file
% set second argument of \begin to the number of references
% (used to reserve space for the reference number labels box)

% biography section
% 
% If you have an EPS/PDF photo (graphicx package needed) extra braces are
% needed around the contents of the optional argument to biography to prevent
% the LaTeX parser from getting confused when it sees the complicated
% \includegraphics command within an optional argument. (You could create
% your own custom macro containing the \includegraphics command to make things
% simpler here.)
%\begin{IEEEbiography}[{\includegraphics[width=1in,height=1.25in,clip,keepaspectratio]{mshell}}]{Michael Shell}
% or if you just want to reserve a space for a photo:

\begin{IEEEbiography}[{\includegraphics[width=1in,height=1.25in,clip,keepaspectratio]{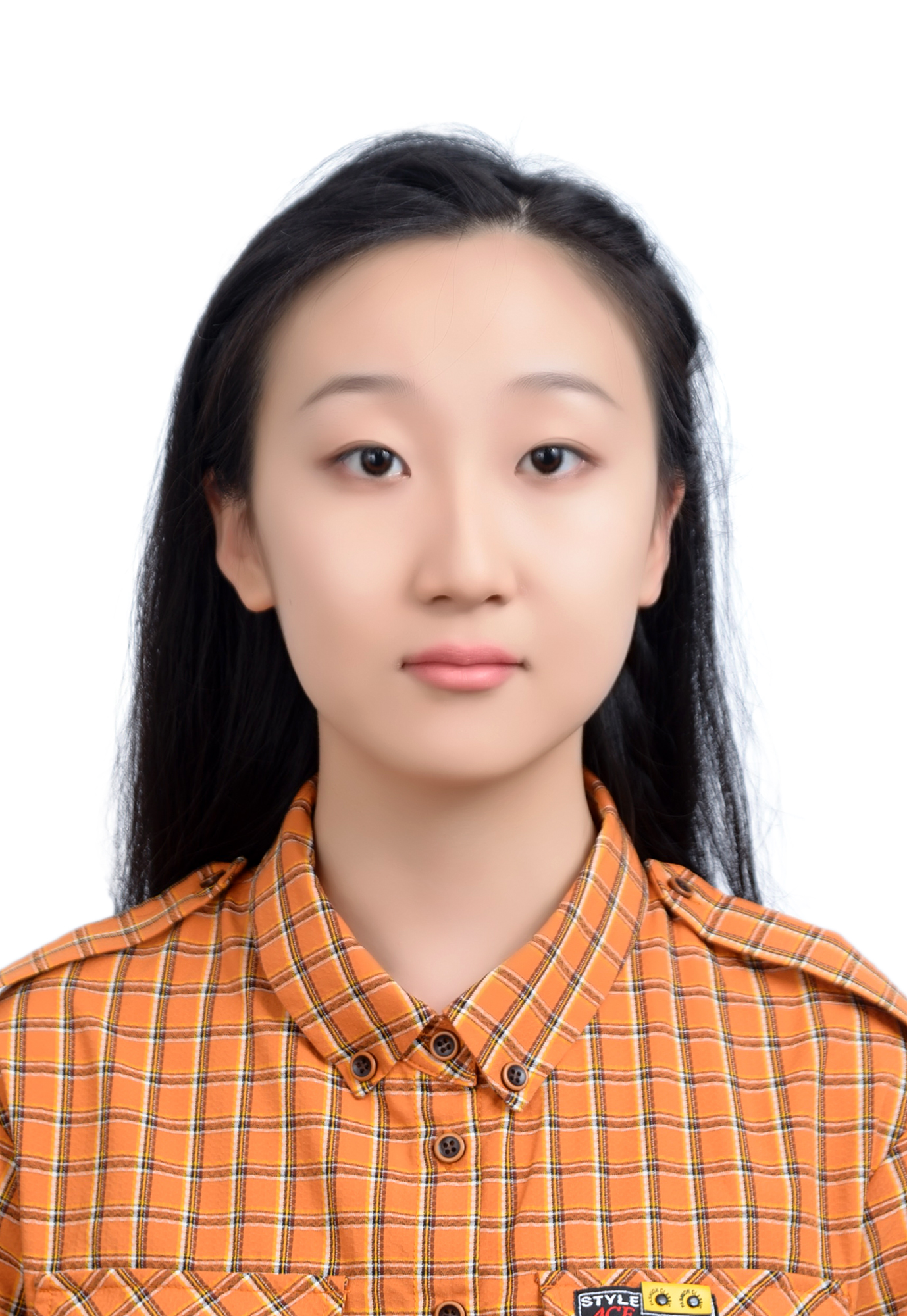}}]
{Zhihang Song} received the B.S. degree in Mechatronic Engineering, Zhejiang University, Hangzhou, China, in 2022. She is currently pursuing the master’s degree in control science and engineering with the Department of Automation, Tsinghua University, China. Her current research interests include multi-sensor data fusion, autonomous driving, and cooperative driving.
\end{IEEEbiography}

\begin{IEEEbiography}[{\includegraphics[width=1in,height=1.25in,clip,keepaspectratio]{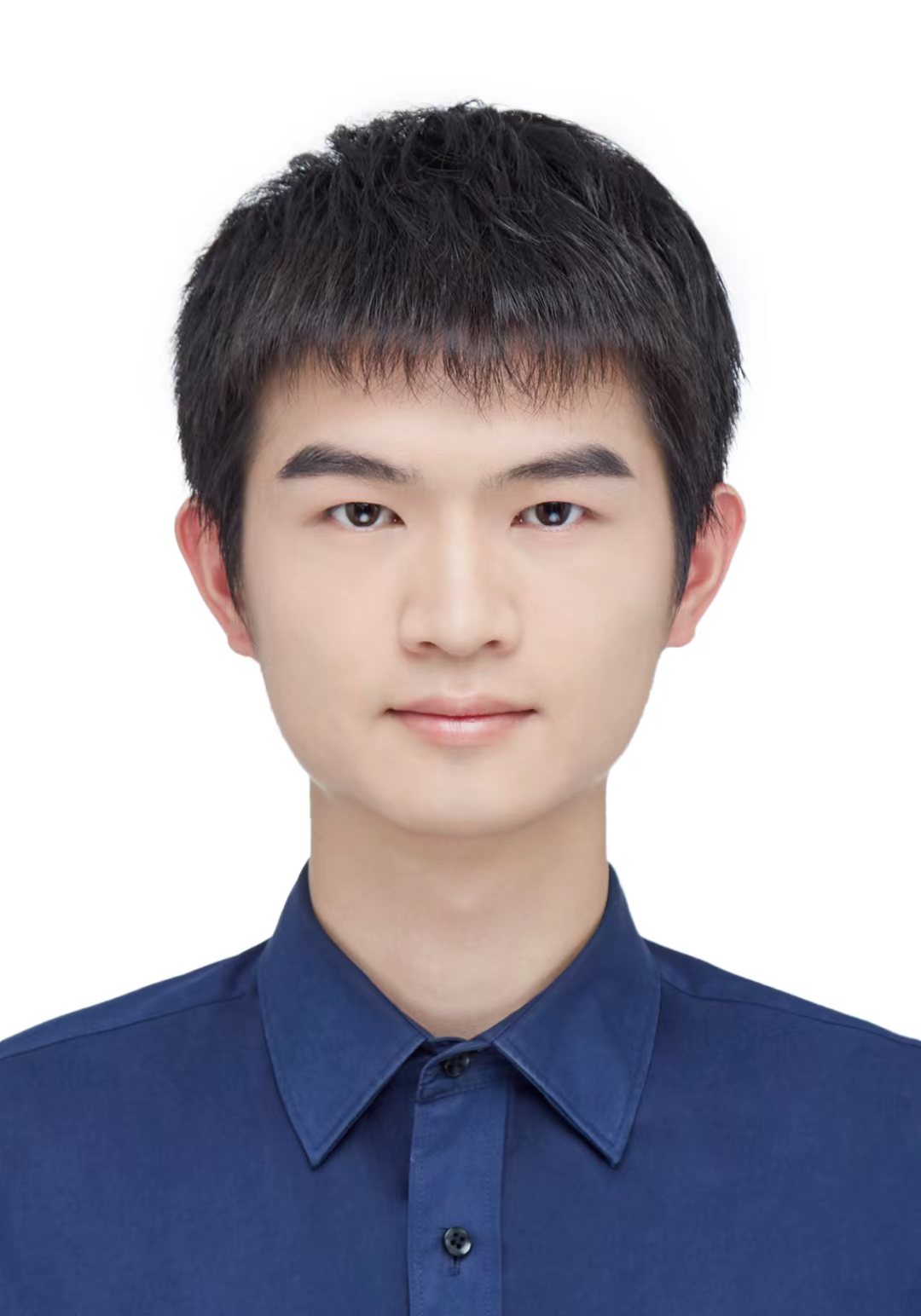}}]
{Zimin He} received the B.S. degree from the Department of Automation, Tsinghua University, Beijing, China, in 2022. He is currently pursuing the master’s degree in control science and engineering with the Department of Automation, Tsinghua University, China. His current research interests include intelligence testing, autonomous driving, and cooperative driving.
\end{IEEEbiography}

\begin{IEEEbiography}[{\includegraphics[width=1in,height=1.25in,clip,keepaspectratio]{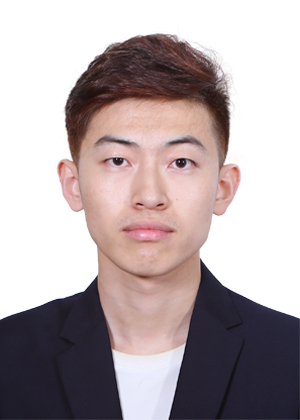}}]
{Xingyu Li} received the B.S. degree from the Department of Automation, Tsinghua University, Beijing, China, in 2022. He is currently pursuing the master’s degree in electronic information with the Department of Automation, Tsinghua University, China. His current research interests include multi-sensor object perception, autonomous driving, and cooperative driving.
\end{IEEEbiography}

\begin{IEEEbiography}[{\includegraphics[width=1in,height=1.25in,clip,keepaspectratio]{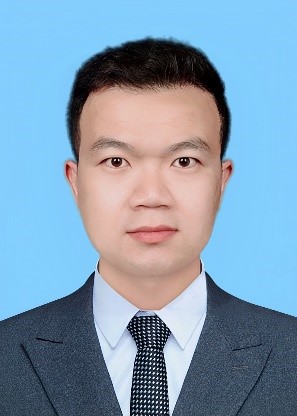}}]
{Qiming Ma} received the B.S. degree from Tongji University, Shanghai, China, in 2018, and Master’s degree in automation from Tsinghua University, Beijing, China, in 2022, where he is currently pursuing the doctoral degree in the Department of Automation. His research interests include intelligent system testing and verification, image processing, and autonomous driving.
\end{IEEEbiography}

\begin{IEEEbiography}[{\includegraphics[width=1in,height=1.25in,clip,keepaspectratio]{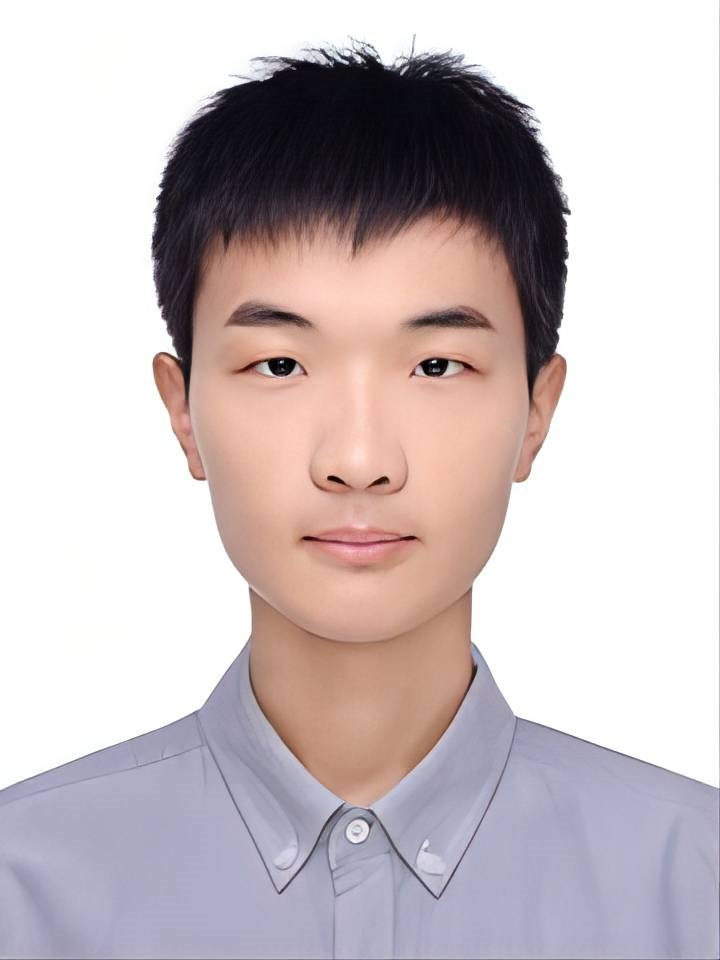}}]
{Ruibo Ming} received the B.S. degree from the Department of Computer Science, Beijing Institute of Technology, Beijing, China, in 2022. He is currently pursuing the master’s degree in electronic information with the Department of Automation, Tsinghua University, Beijing, China. His current research interests include autonomous driving, large language model, and 3D reconstruction.
\end{IEEEbiography}

\begin{IEEEbiography}[{\includegraphics[width=1in,height=1.25in,clip,keepaspectratio]{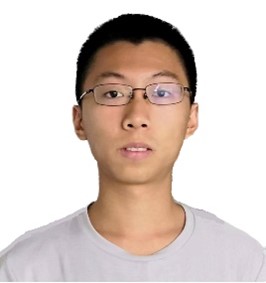}}]
{Zhiqi Mao} received the B.S. degree from the Department of Electronic Engineering, Tsinghua University, Beijing, China, in 2020 and received the Master degree from the Department of Automation, Tsinghua University, Beijing, China. His current research interests include reinforcement learning, autonomous driving, and operation research.
\end{IEEEbiography}

% insert where needed to balance the two columns on the last page with
% biographies
%\newpage

\begin{IEEEbiography}[{\includegraphics[width=1in,height=1.25in,clip,keepaspectratio]{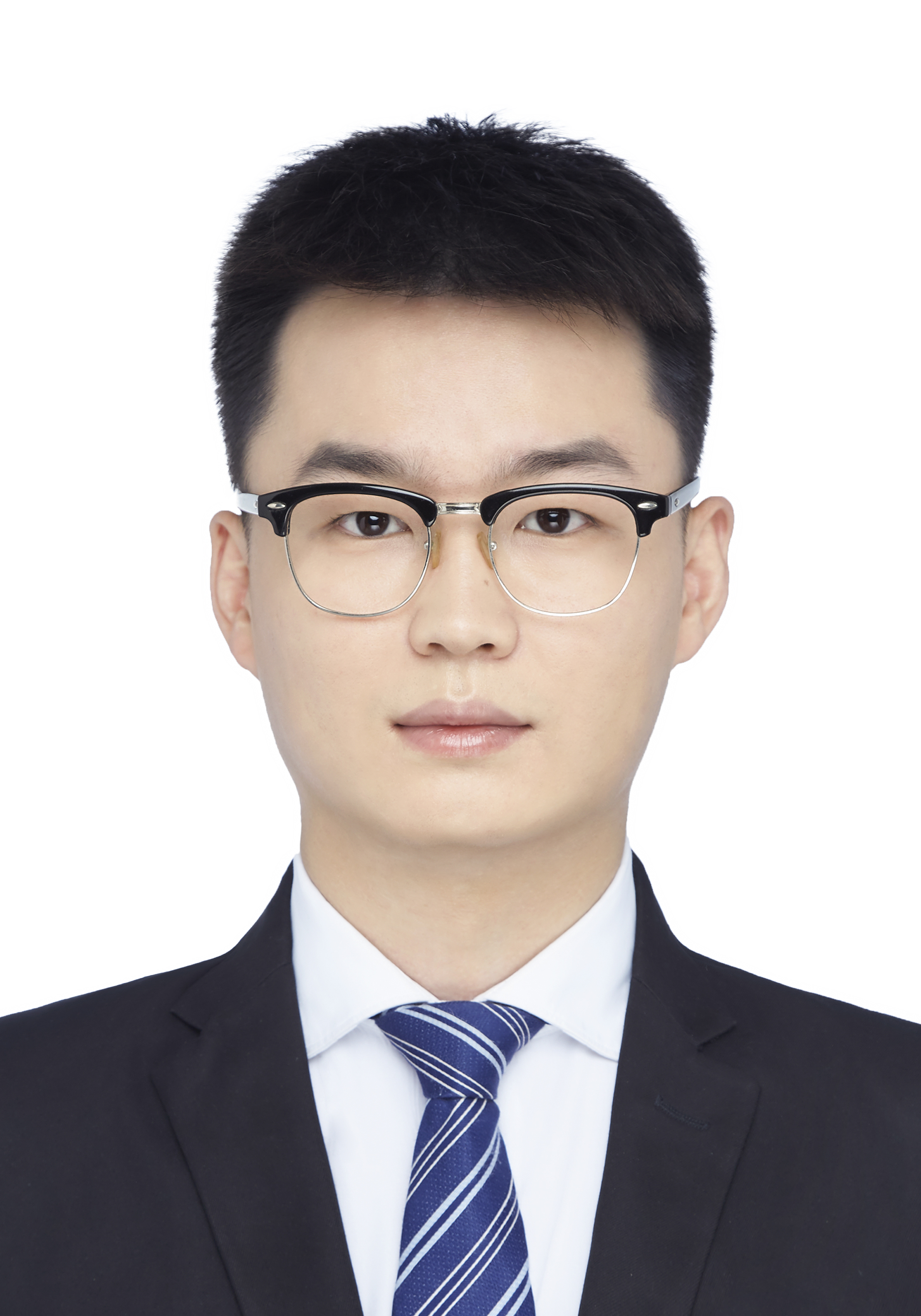}}]
{Huaxin Pei} received the B.S. degree from University of Electronic Science and Technology of China, China, in 2017, and earned the Ph.D. degree in control science and engineering with the Department of Automation, Tsinghua University, China, in 2023. His current research interests include intelligence testing, autonomous driving, cooperative driving, and intelligent vehicle-infrastructure cooperative systems.
\end{IEEEbiography}

\begin{IEEEbiography}[{\includegraphics[width=1in,height=1.25in,clip,keepaspectratio]{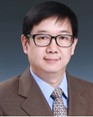}}]
{Lihui Peng} (M’11–SM’15) received the B.Eng., M.Sc., and Ph.D. degrees in measurement science and electronics from Tsinghua University, Beijing, China, in 1990, 1995, and 1998 respectively. In 1990, he started his academic career as a Lecturer with Tsinghua University. He is currently a Professor with the Department of Automation, Institute of Measurement and Applied Electronics, Tsinghua University. He has been a member of three Chinese national technical committees since 2001. He has published more than 100 research articles. His current research interests include multi-sensor data fusion, intelligent sensing and measuring, machine learning based data analytics for measurement and instrumentation, the measurement techniques for multiphase flow, process tomography, flow measurement and instrumentation. Dr. Peng is an Associate Editor of the IEEE TRANSACTIONS ON INSTRUMENTATION AND MEASUREMENT.
\end{IEEEbiography}

\begin{IEEEbiography}[{\includegraphics[width=1in,height=1.25in,clip,keepaspectratio]{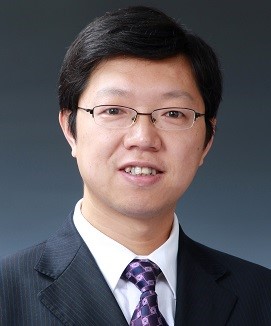}}]
{Jianming Hu} (M’04) received the B.E., M.E., and Ph.D. degrees in 1995, 1998, and 2001, respectively. He is currently an Associate Professor with the Department of Automation (DA), Tsinghua University. He has presided and participated in more than 20 research projects granted from the Ministry of Science and Technology of China, National Science Foundation of China, and other large companies with more than 30 journal articles and more than 150 conference papers. His recent research interests include networked traffic flow, large-scale traffic information processing, intelligent vehicle infrastructure cooperation systems (V2X or Connected Vehicles), and urban traffic signal control.
\end{IEEEbiography}

\begin{IEEEbiography}[{\includegraphics[width=1in,height=1.25in,clip,keepaspectratio]{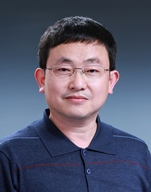}}]
{Danya Yao} (M’04) received the B.S., M.S., and Ph.D. degrees from Tsinghua University, Beijing, China, in 1988, 1990, and 1994, respectively. He is currently a Full Professor with the Department of Automation, Tsinghua University. His research interests include intelligent detection technology, system engineering, mixed traffic flow theory, and intelligent transportation systems.
\end{IEEEbiography}

\begin{IEEEbiography}[{\includegraphics[width=1in,height=1.3in,clip,keepaspectratio]{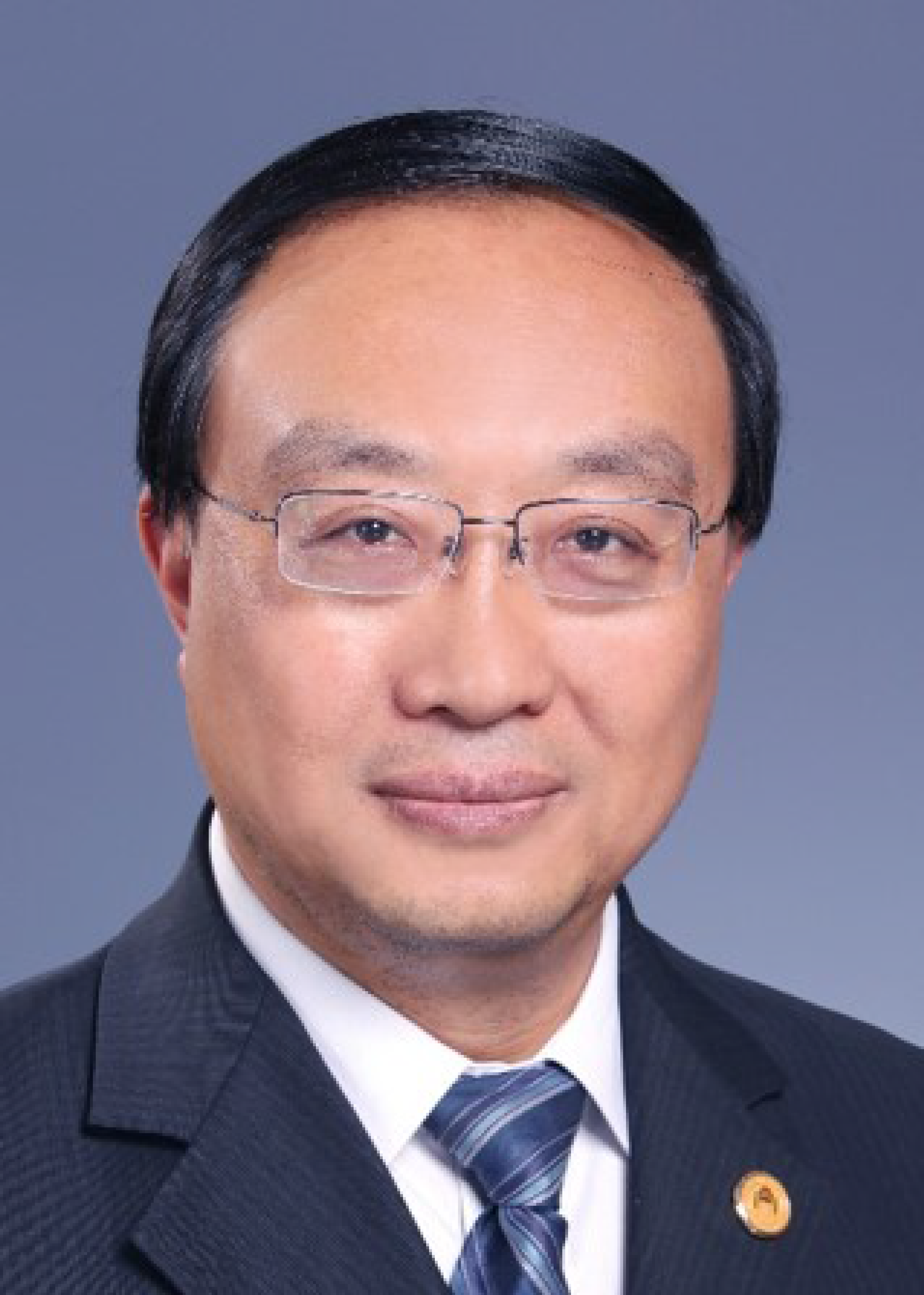}}]{Yi Zhang} (S’01-M’04)
received the B.S. in 1986 and M.S. degree in 1988 from Tsinghua University in China, and earned the Ph.D. degree in 1995 from the University of Strathclyde in the UK. He is a professor in the control science and engineering at Tsinghua University with his current research interests focusing on intelligent transportation systems. His active research areas include intelligent vehicle-infrastructure cooperative systems, analysis of urban transportation systems, urban road network management, traffic data fusion and dissemination, and urban traffic control and management. His research fields also cover the advanced control theory and applications, advanced detection and measurement, systems engineering, etc.
\end{IEEEbiography}

% You can push biographies down or up by placing
% a \vfill before or after them. The appropriate
% use of \vfill depends on what kind of text is
% on the last page and whether or not the columns
% are being equalized.

%\vfill

% Can be used to pull up biographies so that the bottom of the last one
% is flush with the other column.
%\enlargethispage{-5in}

% that's all folks
\end{document}